\documentclass{article}
\usepackage{amsmath,amssymb}
\usepackage{graphicx}%
\newcommand{\Fig}[3]{%
\begin{center}
\parbox{#2cm}{%
\refstepcounter{figure}\includegraphics[width=#2cm]{#1} \noindent {\bf Fig. \thefigure.}\quad
#3}\end{center}}
\newcommand{\FigReg}[4]{%
\begin{center}
\parbox{#2cm}{%
\refstepcounter{figure}\includegraphics[width=#2cm,height=#3cm]{#1} \noindent {\bf Fig. \thefigure.}\quad
#4}\end{center}}
\newcommand{\TwoFig}[4]{%
\begin{flushleft}
\begin{tabular}{lr}
\parbox{7.6cm}{\includegraphics[width=7.6cm]{#1}}  & \parbox{7.6cm}{\includegraphics[width=7.6cm]{#3}} \\
\parbox{7.6cm}{\refstepcounter{figure}{\bf Fig. \thefigure.}\quad #2} & \parbox{7.6cm}{\refstepcounter{figure}{\bf Fig. \thefigure.}\quad #4}\\
\end{tabular}
\end{flushleft}
}

\begin{document}
\begin{center}
{\bf \Large Qualitative and Numerical Analysis of the Cosmological Model Based on a Phantom Scalar field with Self-Action. II. Comparative Analysis of Models of Classical and Phantom Fields.} \\[12pt]
Yu.\,G.~Ignat'ev, A.\,A.~Agathonov\\
N.I. Lobachevsky Institute of Mathematics and Mechanics, Kazan Federal University, \\ Kremleovskaya str., 35, Kazan, 420008, Russia
\end{center}

\begin{abstract}
On the basis of the qualitative analysis and numerical simulation of cosmological models with classical and phantom scalar fields with self-action there have been revealed and refined such models' distinctive features and potential possibilities for their usage as basic models in cosmology.

{\bf keyword} cosmological models, classical and phantom scalar fields, quality analysis, asymptotic behavior, numerical simulation, numerical gravitation
\end{abstract}

This work was funded by the subsidy allocated to Kazan Federal University for the state assignment in the sphere of scientific activities.

\begin{flushleft}
{\bf{Introduction}}
\end{flushleft}

Phantom fields were introduced in gravitation as one of the possible models of the scalar field in 1983 in Author's work \cite{Ignat83_1}. In this article, as well as in later ones (see e.g., \cite{Ignat_Kuz84}, \cite{Ignat_Mif06}) phantom fields were classified as scalar fields with attraction of like-charged particles and were emphasized by means of factor  $\epsilon=-1$ in the energy-momentum tensor of the scalar field\footnote{Let us notice that phantom fields in the context of wormholes and so-called black universes were considered in \cite{Bron1}, \cite{Bron2}.}. Later papers since 2012, were developing the nonminimum theory of scalar interaction on the basis of fundamental scalar charge's concept for both classical and phantom fields \cite{Ignat_12_1_Iz},
\cite{Ignat_12_2_Iz}, \cite{Ignat_12_3_Iz}, \cite{Ignat_13_Mono}. In particular, certain features of phantom fields like peculiarities of inter-particle interaction, were revealed in these works. Since 2014 \cite{Ignat_14_1_stfi}, \cite{Ignat_Dima14_2_GC}, \cite{Ignat_Dima14_2_GC}, \cite{Ignat_Agaf_Dima14_3_GC}, \cite{Ignat_15_1_GC}, \cite{Ignat_Agaf15_2_GC} these researches were enriched to expand the theory of scalar and in particular, phantom fields, to the range of negative masses of particles, degenerated Fermi-system, conformally invariant interactions and suchlike. In 2015, the constructed mathematical models of scalar fields were applied to investigation of the cosmological evolution of systems of interacting particles and scalar fields of both classical and phantom types \cite{Ignat_Mih15_2_Iz}, \cite{Ignat_Agaf_Mih_Dima15_3_AST}, \cite{Ignat_Agaf_16_3}. These researches revealed the unique properties of cosmological evolution of plasma with inter-particle phantom scalar interaction such as existence of giant bursts of the cosmological acceleration, presence of plateau with constant acceleration and suchlike.

However, the constructed numerical models are not able to satisfy a theoretical physicist since they provide no possibility neither for analytical description of such phenomena nor for uncovering the nature of the revealed peculiarities. Due to this there arises a need for qualitative investigation of the cosmological models based on scalar interaction. As a first step, it is required to carry out a qualitative research of the cosmological models, founded on free scalar fields. Such researches for a classical scalar massive scalar field have been carried out since 1985 \cite{Belinsky}, \cite{Zeld}, \cite{Mex1}, \cite{Zhur_01}, \cite{Mex2} (see also \cite{Bron}). In Zhuravlev's work \cite{Zhur_01} the methods of qualitative theory of dynamic systems were used to investigate the two-component cosmological model with minimum interaction. In the Author's paper \cite{Ignat_16_1_stfi} an incorrectness of so-called approach of <<slow rolling>> was shown; having reduced the problem to investigation of the dynamic system on a flat, it has been over again carried out a qualitative and numerical analysis of the standard cosmological model based on classical scalar field. It was also shown that the invariant cosmological acceleration on late stages of expansion has a microscopic oscillating character.
Then the results were generalized for cosmological models with a $\Lambda$ - term \cite{Ignat_16_2_stfi}, \cite{Ignat_17_1_GC}, and in this case the Author managed to confirm the conservation of oscillating character of invariant cosmological acceleration at sufficiently small values of the cosmological term. The method of investigation suggested in cited works, was used in \cite{Zhur_16} and applied to two-component system <<scalar field+liquid>> with an arbitrary potential function $V(\phi)$\footnote{Particularly,for the Higgs potential.}. We will later on return to certain results of this work.

Then, the qualitative analysis of the cosmological model, based on phantom scalar field with self-action was partially carried out  in \cite{Ignat_16_5_Iz}, \cite{Ignat_Agaf_16_6_stfi} and \cite{Ignat_Agaf_2017_2_GC}. In this article we develop and elaborate the results of investigations of the cosmological models, based on classical and phantom scalar fields.


\section{Basic Relations of the Cosmological Model with a Scalar Field}
\vspace{0pt}
\subsection{Equations of the Free Scalar Field with Self-Action}

In cited papers \cite{Ignat_16_5_Iz}, \cite{Ignat_Agaf_16_6_stfi} the Lagrange function of a scalar field with mass $m$ and self-action was written in the following form (see e.g. \cite{Ignat_12_3_Iz}):
\begin{equation} \label{GrindEQ__1_}
L=\frac{\epsilon_1}{8\pi } \left(g^{ik} \Phi _{,i} \Phi _{,k} - \epsilon_2 m^{2} \Phi ^{2} +\frac{\alpha }{2} \Phi ^{4} \right),
\end{equation}
where $\alpha $ is a constant of self-action; for a field with repulsion of like-charged particles $\epsilon_1=1$, for a field with attraction of like-charged particles $\epsilon_1=-1$; for a classical scalar field $\epsilon_2=1$, for a phantom scalar field $\epsilon_2=-1$. Let us notice, that {\it a single} classical scalar field with attraction of like-charged particles can not exist since its energy is strictly negative. The energy-momentum tensor relative to the Lagrange function (\ref{GrindEQ__1_}) is equal to:
\begin{eqnarray} \label{GrindEQ__2_}
T^{ik} =\frac{\epsilon_1}{8\pi } \left(2\Phi ^{,i} \Phi ^{,k} -g^{ik} \Phi _{,j} \Phi ^{,j}+g^{ik} \epsilon_2 m^{2} \Phi ^{2} -g^{ik} \frac{\alpha }{2} \Phi ^{4} \right).
\end{eqnarray}
To reduce the Lagrange function (\ref{GrindEQ__1_}) to denotations that are standard for cosmological models with scalar fields, let us re-write the function in a different normalization, taking into account the fact that an arbitrary constant can be appended to the Lagrange function:
\begin{equation}\label{Lagrange}
L=\frac{\epsilon_1}{8\pi}\left(g^{ik}\phi_{,i}\Phi_{,k}-2V(\Phi)\right),
\end{equation}
where
\begin{eqnarray}
\label{V(phi)}
V(\Phi)=-\frac{1}{4}(\alpha\Phi^4-2\epsilon_2 m^2\Phi^2)\Rightarrow -\frac{\alpha}{4}\left(\Phi^2-\epsilon_2\frac{m^2}{\alpha}\right)^2.
\end{eqnarray}
It is obvious, that such a renormalization of the potential in the Einstein equations will be equivalent to renormalization of the cosmological constant
\begin{equation}\label{Lambda->Lambda}
\Lambda=\Lambda_0-\epsilon_1 \frac{m^4}{2\alpha}; \quad (\alpha\not\equiv 0),
\end{equation}
where $\Lambda_0$ is a certain <<primeval>> value of the cosmological constant. Thus, in terms of the Lagrangian function (\ref{Lagrange}), potential energy (\ref{V(phi)}), the considered potential function in papers  \cite{Ignat_16_5_Iz}, \cite{Ignat_Agaf_16_6_stfi} is equivalent to the Higgs potential (\ref{V(phi)}). In terms of these values the model is completely defined by signs $\epsilon_1$, $\epsilon_2$ and $\alpha$.

Tensor of energy-momentum of a scalar field in terms of these values takes the standard form:
\begin{equation}\label{T_ik}
T_{ik}=\frac{\epsilon_1}{8\pi}\bigl(2\Phi_{,i}\Phi_{,k}-g_{ik}\Phi_{,j}\Phi^{,j}+2V(\Phi)g_{ik}\bigr).
\end{equation}

Equating covariant divergence of this tensor to null, we obtain the equation of a free scalar field:
\begin{equation} \label{GrindEQ__3_}
\square \Phi +V'(\Phi) =0.
\end{equation}
In particular, using  the function of the potential energy in form  (\ref{V(phi)}) we obtain from~(\ref{GrindEQ__3_}):
\begin{equation} \label{Eq_fild0}
\Box\Phi+m^2_*\Phi=0,
\end{equation}
where $m_*$ is an effective mass of a scalar boson
\begin{equation}
\label{m_*}
m^2_* \equiv \epsilon_2 m^2-\alpha\Phi^2,
\end{equation}
which can be also an imaginary value.

Let us also write out the Einstein equations with the cosmological  $\Lambda >0$ term\footnote{ We use Plank system of units:  $G=c=\hbar =1$ . }
\begin{equation} \label{GrindEQ__5_}
R^{ik} -\frac{1}{2} Rg^{ik} =\Lambda g^{ik} +8\pi T^{ik} ,
\end{equation}
where it is necessary to take into account the relation between the  <<primeval>> value of the cosmological term and its effective cross-section (\ref{Lambda->Lambda}).

\subsection{Self-Consistent Equations for the Space - Flat Friedman Model}
Let us write out self-consistent equations of the space - flat cosmological  \linebreak model
\begin{equation} \label{GrindEQ__6_}
ds^{2} =dt^{2} -a^{2} (t)(dx^{2} +dy^{2} +dz^{2})
\end{equation}
-- the Einstein equation
\begin{equation} \label{GrindEQ__7_}
3\frac{\dot{a}^{2} }{a^{2} } =\epsilon_1 \left(\dot{\Phi }^{2} + \epsilon_2 m^{2} \Phi ^{2} -\frac{\alpha }{2} \Phi ^{4} \right) +\Lambda
\end{equation}
and the equation of massive scalar field with cubic nonlinearity \footnote{ Here and further it is $\dot{f}\equiv df/dt$.}:
\begin{equation} \label{GrindEQ__8_}
\ddot{\Phi }+3\frac{\dot{a}}{a} \dot{\Phi } + \epsilon_2 m_{*}^{2} \Phi =0.
\end{equation}
In this case the energy - momentum tensor \eqref{GrindEQ__2_} has a structure of energy -- momentum tensor of isotropic liquid with the following energy density and pressure:
\begin{eqnarray} \label{GrindEQ__9_}
\varepsilon =\frac{\epsilon_1}{8\pi } \left(\dot{\Phi }^{2} + \epsilon_2 m^{2} \Phi ^{2} -\frac{\alpha }{2} \Phi ^{4} \right);\nonumber\\
p=\frac{\epsilon_1}{8\pi } \left(\dot{\Phi }^{2} -\epsilon_2 m^{2} \Phi ^{2} +\frac{\alpha }{2} \Phi ^{4} \right),
\end{eqnarray}
so that:
\[\varepsilon +p=\frac{\epsilon_1\dot{\Phi}^{2}}{4\pi}.\]

\subsection{Kinematic Invariants}
Further, we will need also the values of two kinematic invariants of the Friedman Universe:
\begin{equation} \label{GrindEQ__10_}
H(t)=\frac{\dot{a}}{a} \ge 0;{\rm \; \; }\Omega (t)=\frac{a\ddot{a}}{\dot{a}^{2} } \equiv 1+\frac{\dot{H}}{H^{2} }
\end{equation}
the Hubble constant and the invariant cosmological acceleration.

\section{A Qualitative Analysis}

\subsection{Reducing the System of Equations to the Canonical Form}
Let us reduce the field equation \eqref{GrindEQ__8_} to the form of normal autonomous system of ordinary differential equations in the plane ~$\{ \Phi ,Z\} $ using the fact that the Hubble constant can be expressed from the Einstein equation \eqref{GrindEQ__7_} through functions $\Phi ,{\rm \; }\dot{\Phi }$,  proceeding to dimensionless Compton time:
\[mt=\tau;\quad (m\not\equiv0) \]
and carrying out a standard change of variables $\Phi '=Z(t)$:
\begin{eqnarray} \label{GrindEQ__11_}
\Phi ' &=& Z;\nonumber\\
Z' &=& \displaystyle-\sqrt{3} \sqrt{ \epsilon_1 \left(Z^{2} + \epsilon_2 \Phi ^{2} -\frac{\alpha _{m} }{2} \Phi ^{4} \right) + \Lambda _{m}} \;Z \displaystyle - \epsilon_2 \Phi +\alpha _{m} \Phi ^{3} ,
\end{eqnarray}
where $f'\equiv df/d\tau $ and the following denotations are introduced:

\[\Lambda _{m} \equiv \frac{\Lambda }{m^{2} } ;{\rm \; \; }\alpha _{m} \equiv \frac{\alpha }{m^{2} } . \]
In this case:

\begin{equation} \label{GrindEQ__12_}
H=m\frac{a'}{a} \equiv mh;{\rm \; \; \; }\Omega =\frac{aa''}{a'^{2} } \equiv 1+\frac{h'}{h^{2} } .
\end{equation}

Thus, we have an autonomous two-dimensional system in the phase plane $\{ \Phi ,Z\} $. To reduce it to standard denotations of the qualitative theory of differential equations (see e.g., [15]) let us put the following:
\begin{eqnarray}\label{GrindEQ__13_}
&&\Phi =x;\;Z=y;\nonumber\\
&&P(x,y)=y;\nonumber\\
&&Q(x,y)=-\sqrt{3} \sqrt{ \epsilon_1 \left(y^{2} + \epsilon_2 x^{2} -\frac{\alpha _{m} }{2} x^{4} \right) + \Lambda _{m}} \;y \displaystyle - \epsilon_2 x +\alpha _{m} x^{3}.
\end{eqnarray}
The corresponding normal system of equations in standard denotations have the next form:
\begin{equation} \label{GrindEQ__14_}
x'=P(x,y);{\rm \; \; \; }y'=Q(x,y) .
\end{equation}
In order the system of differential equations \eqref{GrindEQ__11_} (and (14)) to have a real solution, the following inequality must be satisfied: :
\begin{equation} \label{GrindEQ__15_}
\epsilon_1 \left(y^{2} + \epsilon_2 x^{2} -\frac{\alpha _{m} }{2} x^{4} \right) + \Lambda _{m} \ge 0.
\end{equation}

\subsection{Singular Points of the Dynamic System}
Singular points of the dynamic system are defined by the following equations (see e.g.,  [15]) :
\begin{equation} \label{GrindEQ__16_}
M:{\rm \; \; }P(x,y)=0;{\rm \; }Q(x,y)=0.
\end{equation}
It is obvious, that at any values of $\alpha _{m} $ and $\Lambda _{m} \ge 0$ the system of algebraic equations \eqref{GrindEQ__15_} has an unique non-trivial solution just as in papers [13]-[14]
\begin{equation} \label{GrindEQ__17_}
x=0;y=0{\rm \; }\Rightarrow M_{0} (0,0) .
\end{equation}
Moreover, in case of the same signs of $\epsilon_2$ and $\alpha _{m}$, the following non-trivial symmetrical solutions are possible:
\begin{equation} \label{GrindEQ__18_}
x=x_\pm=\pm \frac{1}{\sqrt{\epsilon_2\alpha_{m}}};y=0\quad \Rightarrow M_\pm(x_\pm,0).
\end{equation}
Substituting the solutions \eqref{GrindEQ__18_} in the condition \eqref{GrindEQ__15_}, we obtain the requirement of solutions in the singular points \eqref{GrindEQ__17_} and \eqref{GrindEQ__18_} to be of real type:
\begin{equation} \label{GrindEQ__19_}
(17)\rightarrow \Lambda_m\ge 0;(18)\rightarrow\Lambda_m +\frac{\epsilon_1}{2\alpha_m}\ge 0.
\end{equation}
%

\subsection{Characteristic Equation and Qualitative Analysis in Case Near the Null Singular Point}
Let us calculate the derivatives of functions  \eqref{GrindEQ__13_} in a null singular point \eqref{GrindEQ__16_} at $\Lambda _{m} \ge 0$ :

\begin{equation}
\begin{array}{ll}
\displaystyle\left.\frac{\partial P}{\partial x} \right|_{M_0} =0;&\hspace{1cm}\displaystyle \left.\frac{\partial P}{\partial y}\right|_{M_0}=1;\\
\displaystyle\left.\frac{\partial Q}{\partial x} \right|_{M_0 }=-\epsilon_2;&\hspace{1cm} \left. \displaystyle\frac{\partial P}{\partial y} \right|_{M_0} =\displaystyle -\sqrt{3\Lambda_{m}}.
\end{array}
\end{equation}
Thus, we obtain a characteristic equation and its roots (see [15]):
\begin{eqnarray} \label{GrindEQ__20_}
\left|
\begin{array}{cc}
{-\lambda } & {1} \\[12pt]
{-\epsilon_2} & {-\lambda -\sqrt{3\Lambda _{m} } }
\end{array}
\right|=0
\quad\Rightarrow\quad \lambda _{\pm } =-\frac{\sqrt{3\Lambda _{m} } }{2} \pm \frac{\sqrt{3\Lambda _{m} - 4\epsilon_2} }{2}.
\end{eqnarray}

\subsection{Numerical Integration of a Model Without Cosmological Constant ($\Lambda=0$) with a Classical Scalar Field ($\epsilon_1=1$, $\epsilon_2=1$) Without a Self-Action ($\alpha=0$) }

In this case the roots of the characteristic equation (\ref{GrindEQ__19_}) take the next values:

\begin{equation}\label{lambda1}
\lambda=\pm i.
\end{equation}
Since the eigenvalues turned to be purely imaginary ones, the unique singular point (\ref{GrindEQ__17_}) of the dynamic system is its center
(see \cite{Ignat_16_2_stfi}). In this case the phase trajectory of the dynamic system at $\tau\to+\infty$ is winded around this center, making an infinite number of turns.
%


\subsection{Numerical Integration of the Model with Cosmological Constant ($\Lambda>0$) with a Classical Scalar Field ($\epsilon_1=1$, $\epsilon_2=1$) Without a Self-Action ($\alpha_m=0$) }

In this case the roots of the characteristic equation (\ref{GrindEQ__19_}) take the following values:

\begin{equation}\label{lambda2}
\lambda _{\pm } =-\frac{\sqrt{3\Lambda _{m} } }{2} \pm \frac{\sqrt{3\Lambda _{m} - 4} }{2}.
\end{equation}
The following cases are possible \cite{Ignat_Agaf_16_6_stfi}:

1) $\Lambda _{m} < 4/3$ --- the eigenvalues complex conjugate of negative real parts -- attractive focus.

2) $\Lambda _{m} > 4/3$ --- the eigenvalues real negatives -- stable nodes.

\TwoFig{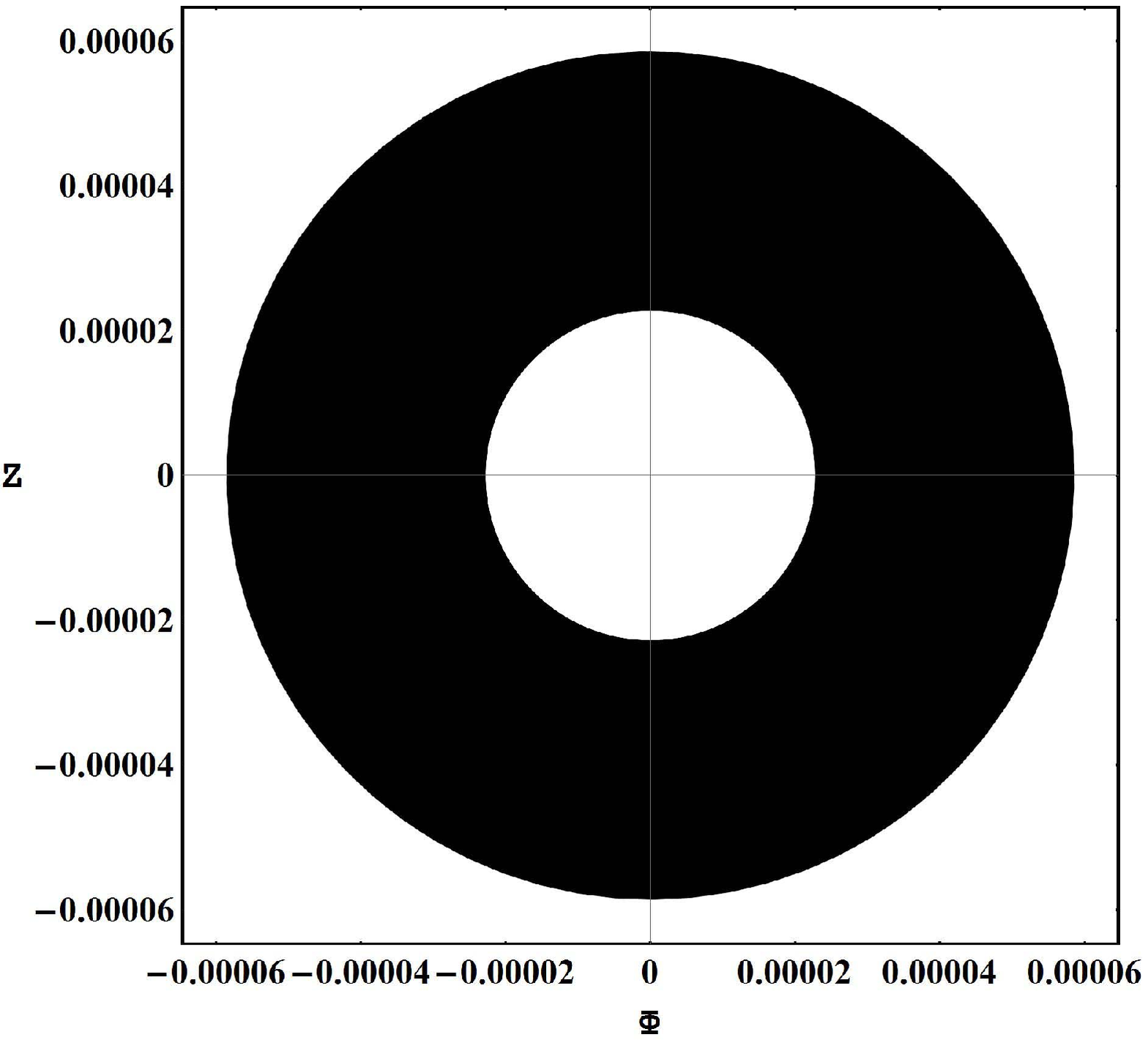}{The phase portrait of the system \eqref{GrindEQ__11_} in a large scale for the case  $\alpha_m=0$; $\Lambda_m=0$; $\Phi(0)=100$; $Z(0)=0$; $\tau=20000\div 50000$.\label{fig::1_1_2}}{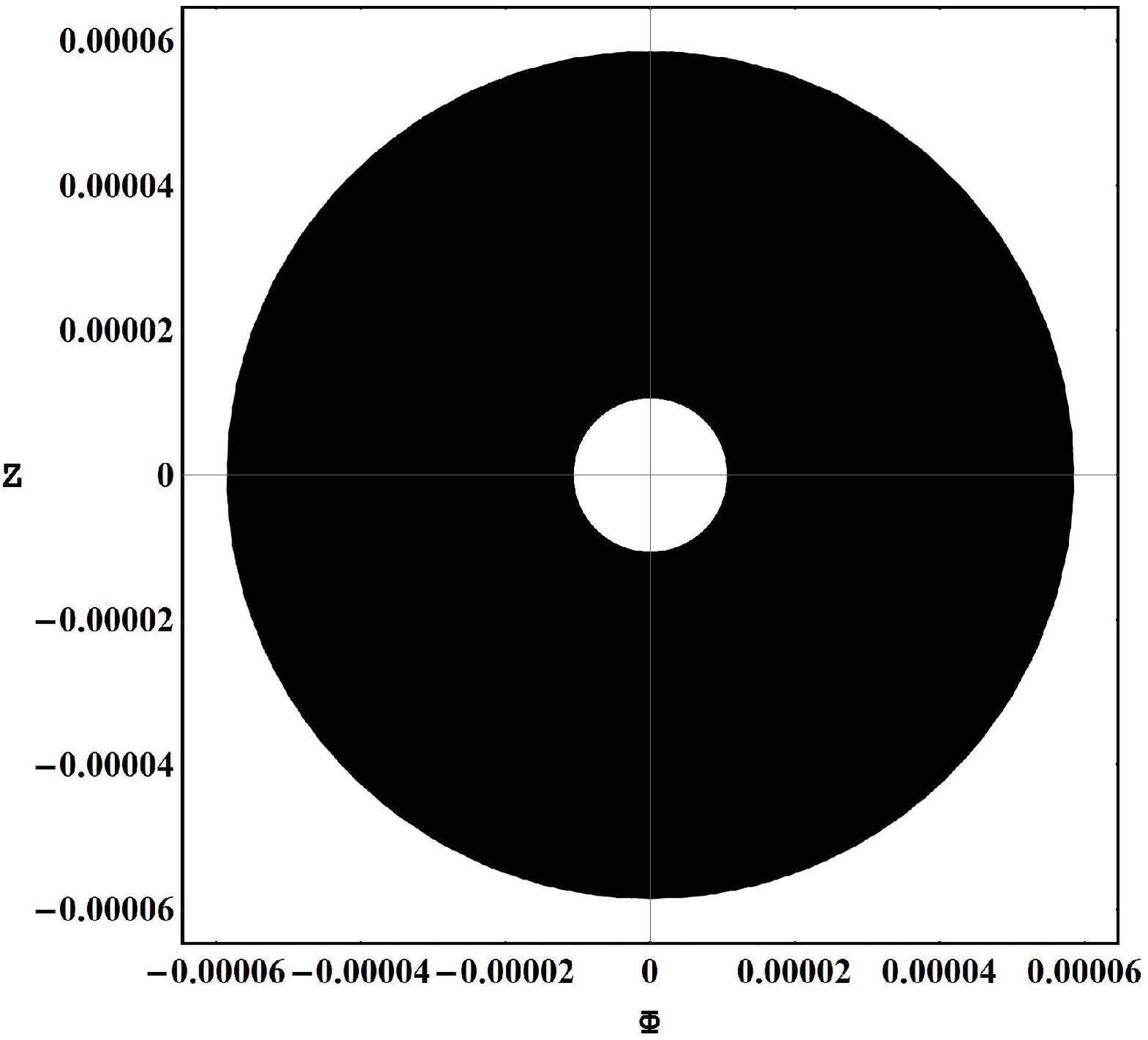}{The same case $\tau=20000\div 105000$.\label{fig::1_1_3}}

It can be shown that the singular point's character does not change when accounting the second order of perturbation theory since all the second partial derivatives of a dynamic system at central point are equal to null:
\begin{equation}\label{secondpartial}
\left.\frac{\partial^2 P}{\partial x^2} \right|_{M_0} = \left.\frac{\partial^2 P}{\partial x \partial y} \right|_{M_0} = \left.\frac{\partial^2 P}{\partial y^2} \right|_{M_0} = \left.\frac{\partial^2 Q}{\partial x^2} \right|_{M_0} = \left.\frac{\partial^2 Q}{\partial x \partial y} \right|_{M_0} = \left.\frac{\partial^2 Q}{\partial y^2} \right|_{M_0} = 0.
\end{equation}

\TwoFig{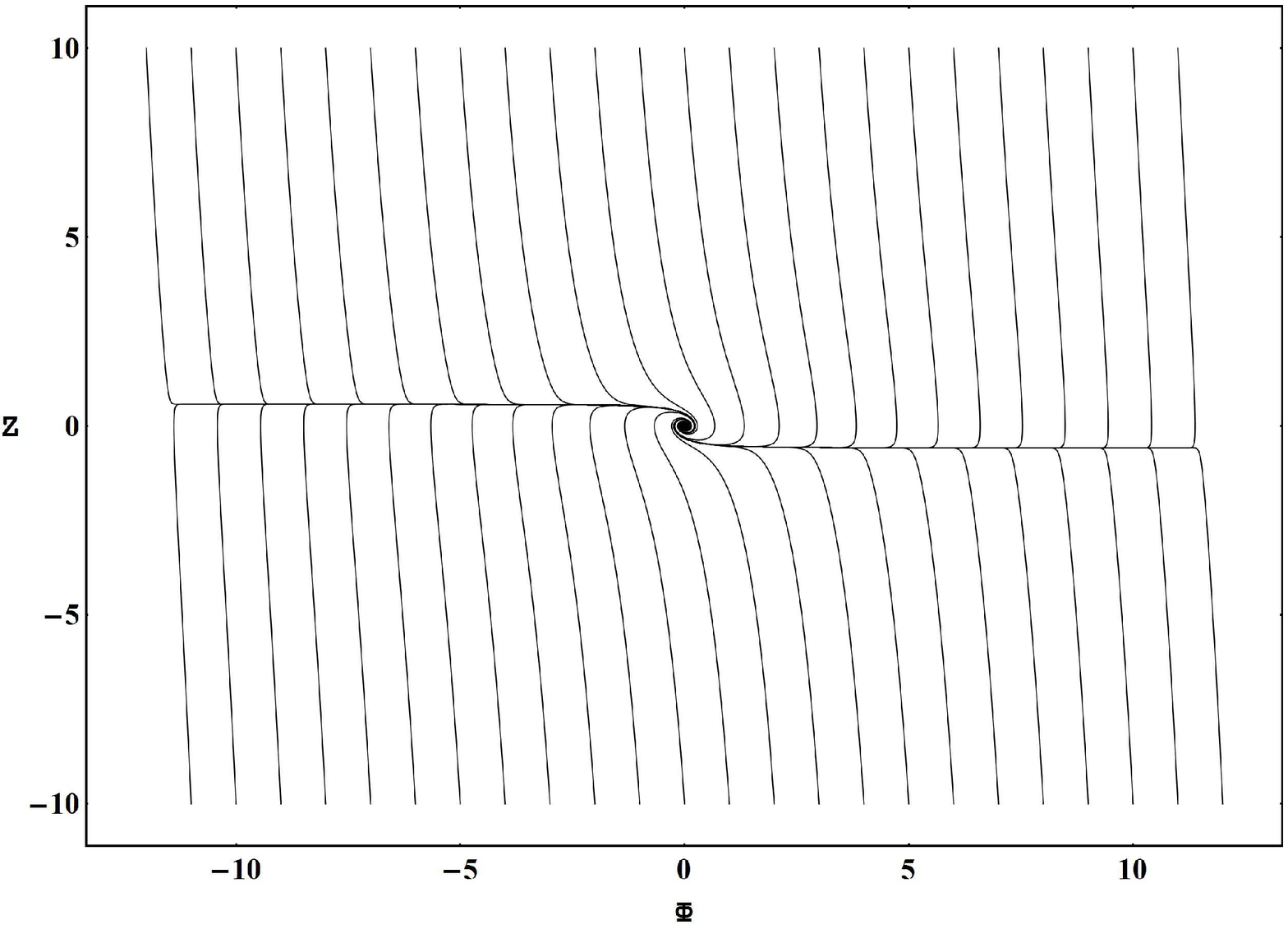}{The phase poratrait of the system \eqref{GrindEQ__11_} in a large scale for the case $\alpha_m=0$;\linebreak $\Lambda_m=0.0001$; $\tau=0 \div 90$.}%
{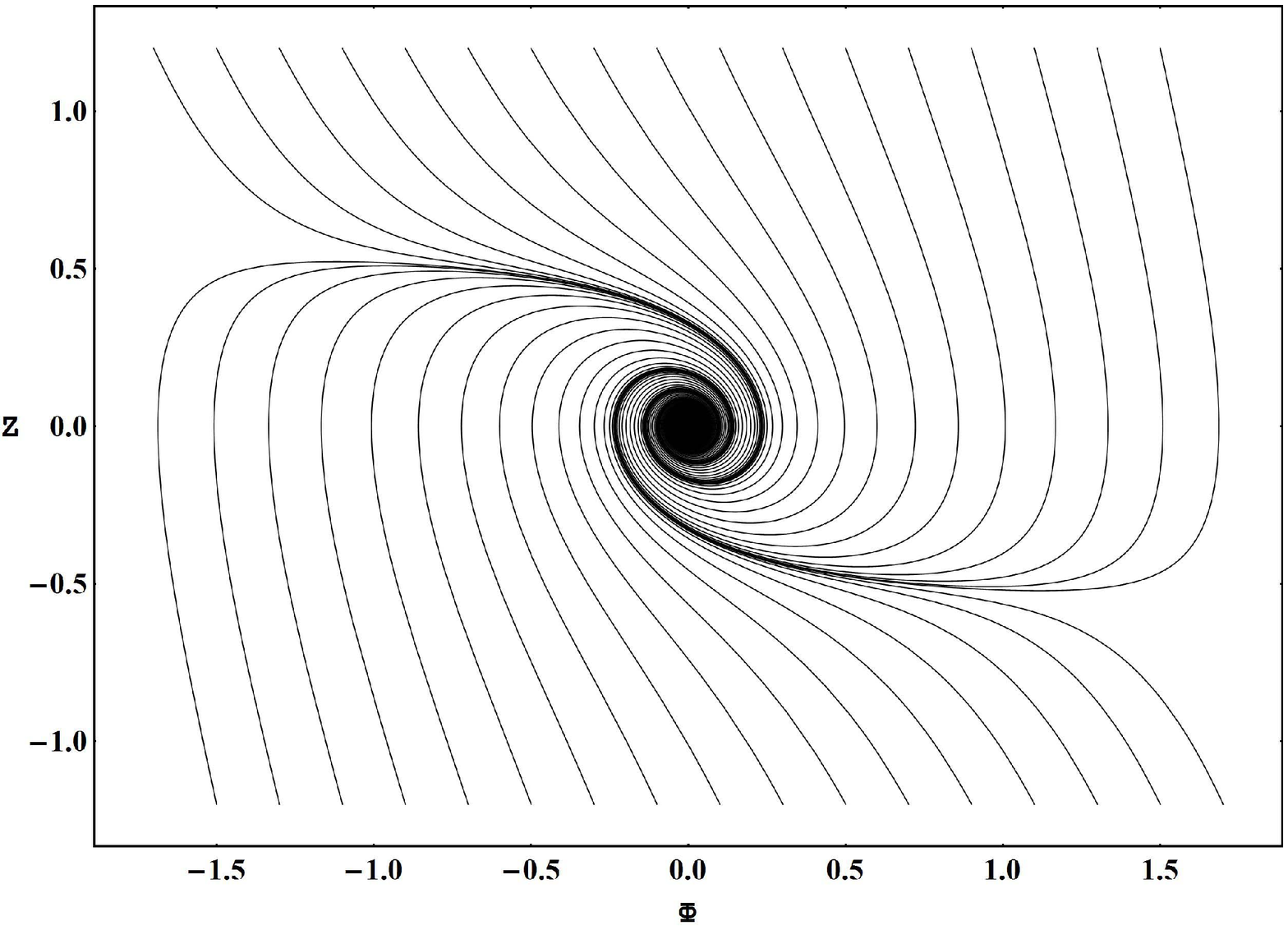}{The same case, neighbourhood of a singular point.}

\subsection{Numerical Integration of the Model with Cosmological Constant ($\Lambda>0$) with a Phantom Scalar Field ($\epsilon_1=-1$, $\epsilon_2=-1$) Without a Self-Action ($\alpha_m=0$) }

In this case the roots of the characteristic equation(\ref{GrindEQ__19_}) take the following values:

\begin{equation}\label{lambda3}
\lambda _{\pm } =-\frac{\sqrt{3\Lambda _{m} } }{2} \pm \frac{\sqrt{3\Lambda _{m} + 4} }{2}.
\end{equation}
The roots of the characteristic equation at any $\Lambda_m$ are the real ones of different signs --- that said the singular point is a saddle.

\subsection{Numerical Integration of a Model Without Cosmological Constant ($\Lambda>0$) with a Phantom Scalar Field ($\epsilon_1=-1$, $\epsilon_2=-1$) with a Self-Action ($\alpha_m < 0$) }

In this case the roots for the central point are a saddle again:

\begin{equation*}\label{lambda2}
\lambda _{\pm } =-\frac{\sqrt{3\Lambda _{m} } }{2} \pm \frac{\sqrt{3\Lambda _{m} + 4} }{2}.
\end{equation*}
\Fig{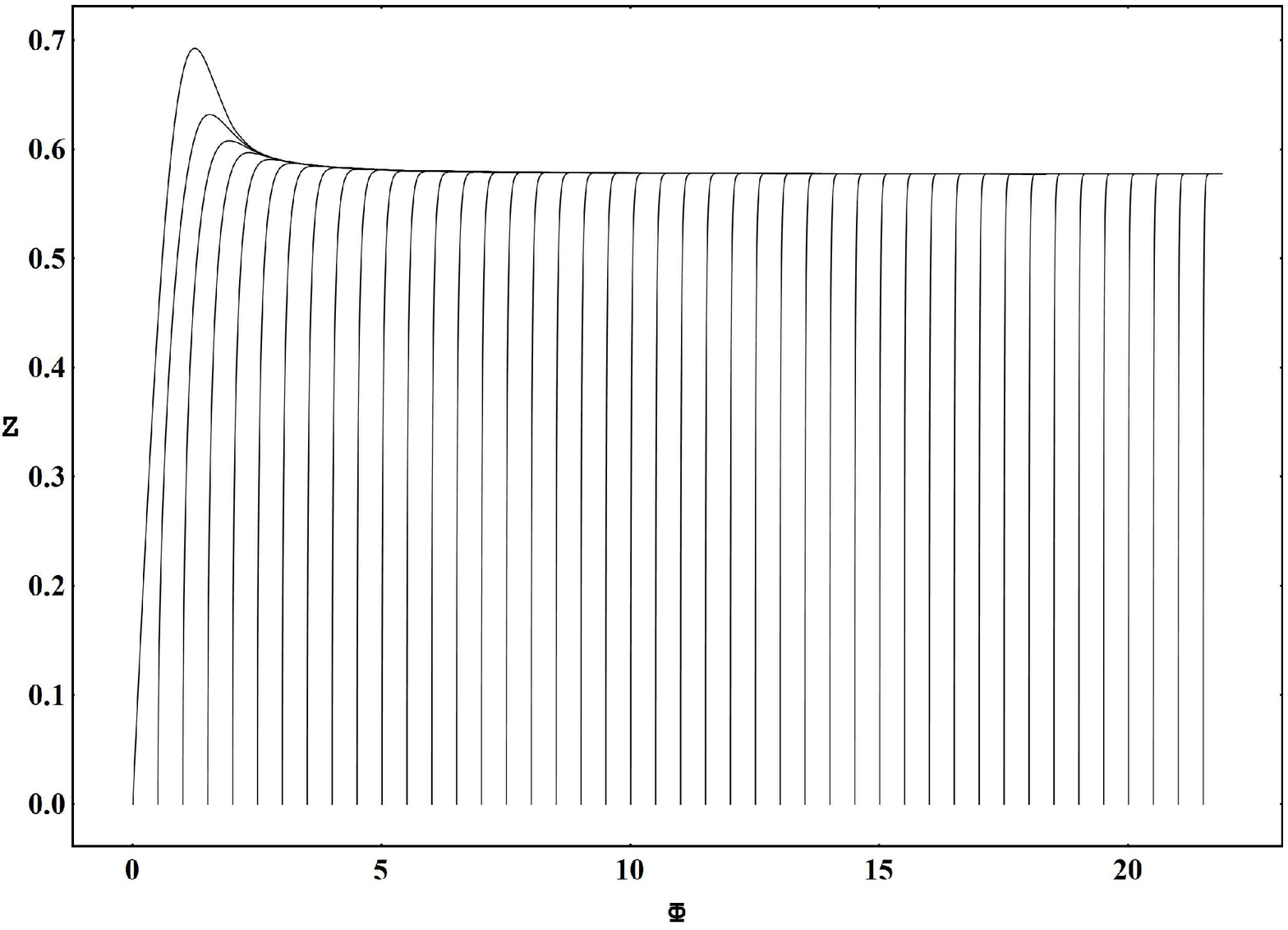}{10}{The phase portrait of the system \eqref{GrindEQ__11_} in a large scale for the case $\alpha_m=0$; $\Lambda_m=0.0001$; $\tau=0 \div 30$.\label{fig::2_2_1}}

The derivatives of functions \eqref{GrindEQ__13_} in the singular points \eqref{GrindEQ__18_} at $\Lambda _{m} >0$ are equal to:
\begin{equation*}
\begin{array}{rl}
\displaystyle\left.\frac{\partial P}{\partial x} \right|_{M_\pm} =0;&\hspace{1cm}\displaystyle \left.\frac{\partial P}{\partial y}\right|_{M_\pm }=1;\\
\displaystyle\left.\frac{\partial Q}{\partial x} \right|_{M_\pm }=-2;&\hspace{1cm} \left. \displaystyle\frac{\partial P}{\partial y} \right|_{M_\pm} =-\displaystyle \sqrt{3}\sqrt{\Lambda _{m}-\frac{1}{2\alpha _{m}}}.
\end{array}
\end{equation*}
Characteristic equations for both singular points coincide and proper points have the same type (see [15]):
\begin{eqnarray} \label{GrindEQ__22_}
\left|\begin{array}{cc}
-\lambda  & 1 \\
1 &\displaystyle -\lambda -\sqrt{3}\sqrt{\Lambda_m -\frac{1}{2\alpha_m}}\\[12pt]
\end{array}\right|=0
\Rightarrow\quad \lambda_\pm =-\frac{\sqrt{3}\sqrt{\Lambda _m-\displaystyle\frac{1}{2\alpha_m}}}{2}
\pm \frac{\sqrt{3}\sqrt{\Lambda_m-\displaystyle\frac{1}{2\alpha_m}-\frac{8}{3}}}{2}.
\end{eqnarray}
In consequence of \eqref{GrindEQ__19_} the radicand in the first term \eqref{GrindEQ__22_} is strictly greater than null, therefore there are next three cases possible:\\
1) $\Lambda_m-1/2\alpha_m-8/3>0$ -- in such case both eigenvalues are real negative ones. In this case the solution contains \textit{two symmetrical attractive (stable) non-degenerated nodes}. All the phase trajectories in the neighbourhood of such singular points come to these points at $t\to\infty$ and apart from two exceptional ones are tangent to an eigenvector of minimal length.

\noindent
2) $\Lambda_m-1/2\alpha_m-8/3=0$ -- in this case both eigenvalues are negative and equal to each other. In this case the solution contains \textit{two symmetrical degenerated nodes}.\\
3) $\Lambda_m-1/2\alpha_m-8/3<0$ -- in this case both eigenvalues are complex conjugate and their real parts are negatives. In this case the solution contains\textit{ two symmetrical attractive focuses}.

In case of two symmetrical focuses it is easy to find a limiting value  $h_\infty$, to which the Hubble constant tends to at $t \to \infty$. Substituting the coordinates of focuses $M_{\pm}(\pm\frac{1}{\sqrt{-\alpha}}, 0)$ into the system (\ref{GrindEQ__11_}), we find:
\begin{equation}
h_\infty = \sqrt{\frac{1}{3}\left( \Lambda_m - \frac{1}{2\alpha_m} \right)}.
\end{equation}

\TwoFig{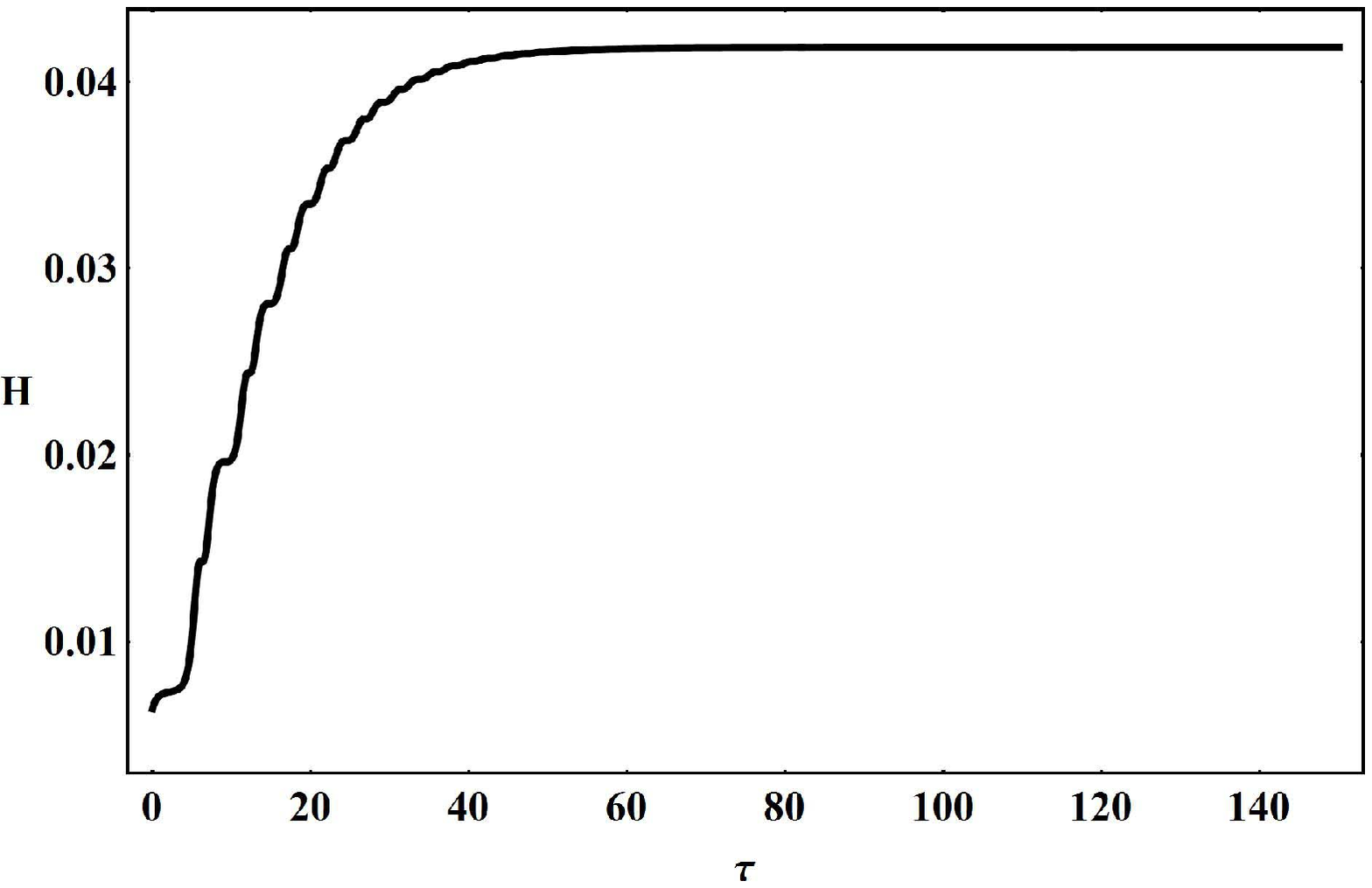}{Evolution of reduced Hubble constant for the case $\alpha_m=-100$; $\Lambda_m=0.00001$; $\Phi(0)=0.4$; $Z(0)=-0.4$; $\tau=0 \div 150$.\label{fig::3_2_2}}%
{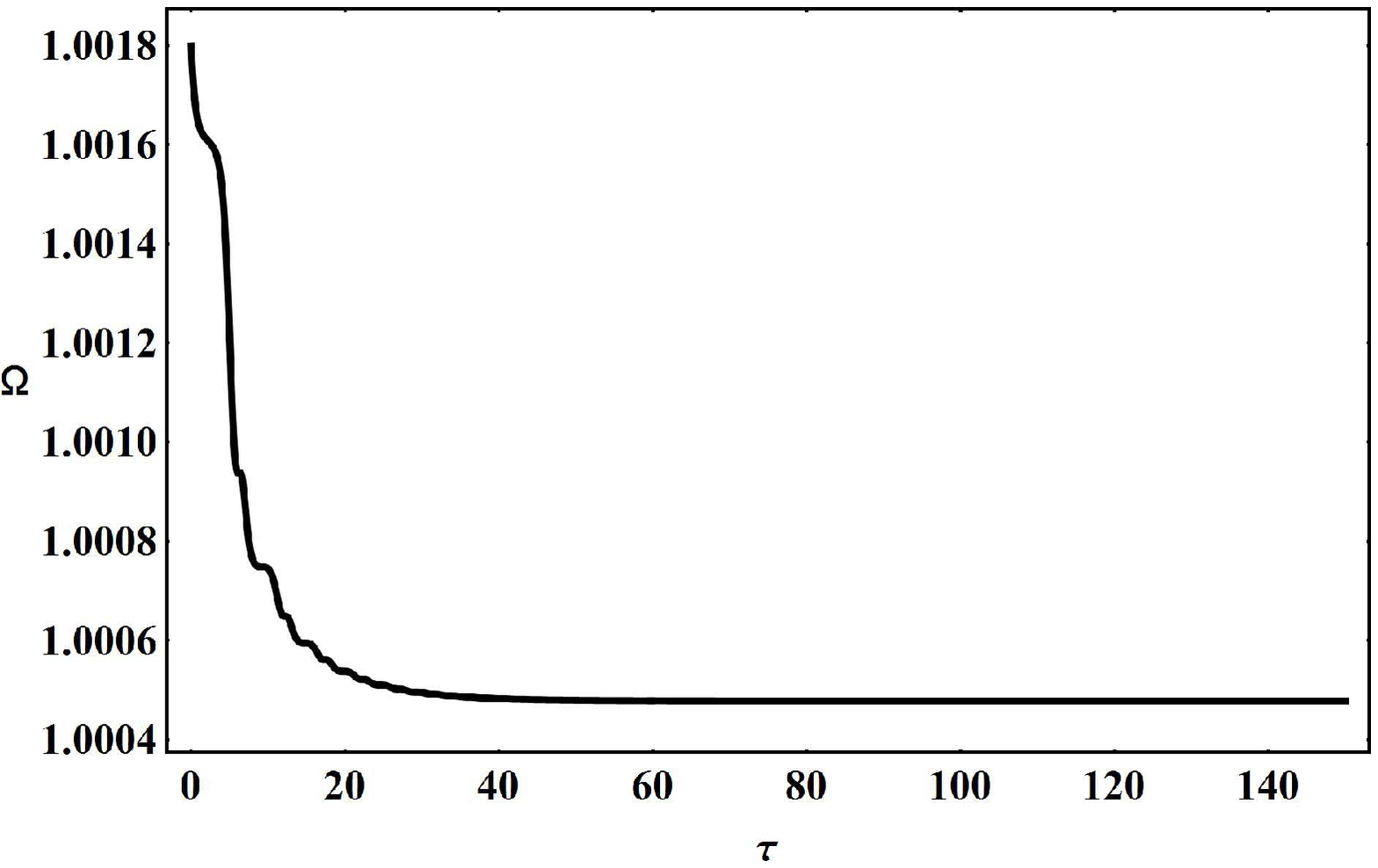}{Evolution of the Cosmological Acceleration for the same case.\label{fig::3_2_3}}
\FigReg{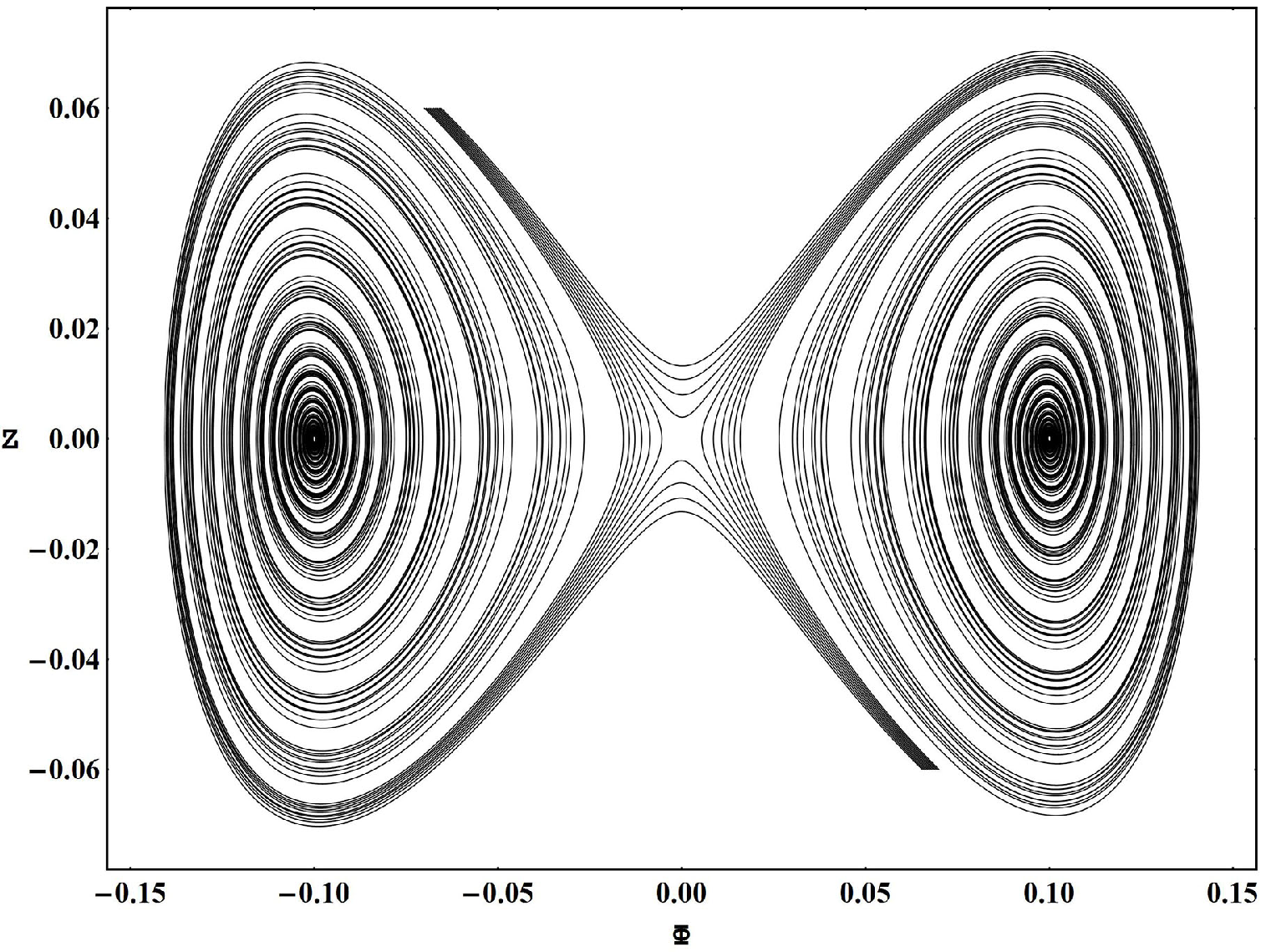}{14}{7}{A phase portrait of the system \eqref{GrindEQ__11_} in a large scale for the case $\alpha_m=-100$; $\Lambda_m=0.00001$;  $\{\Phi(0) = -0.07 + (j - 1)*.0005, j=1..10;
Z0 = 0.06\}$, $\{\Phi(0) = 0.07 + (j - 1)*.0005, j=1..10; Z0 = -0.06\}$; $\tau=0 \div 90$.\label{fig::3_2_1}}

\TwoFig{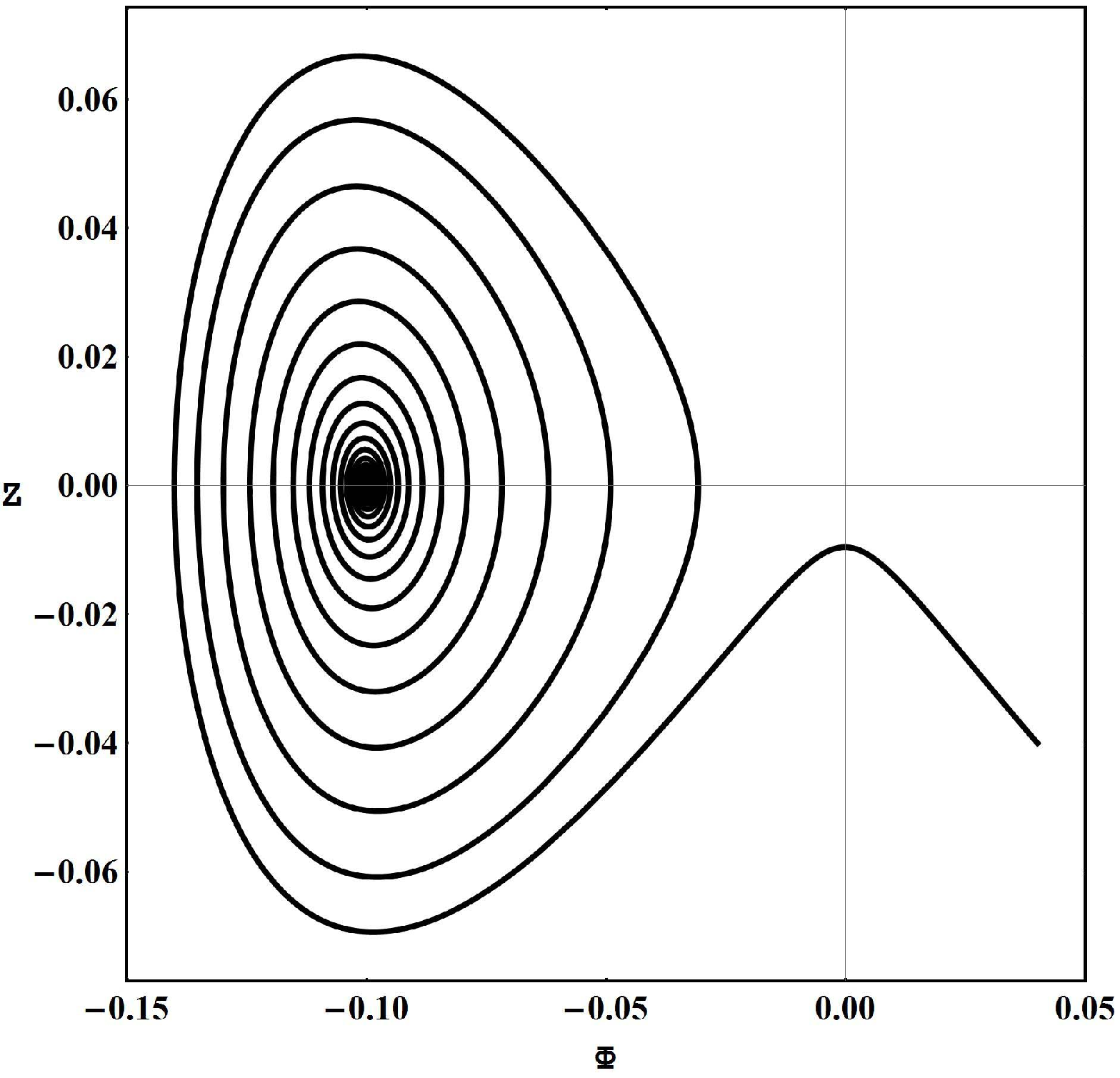}{A phase portrait of the system for the same case, axes $\Phi$, $Z$.\label{fig::3_2_4a}}{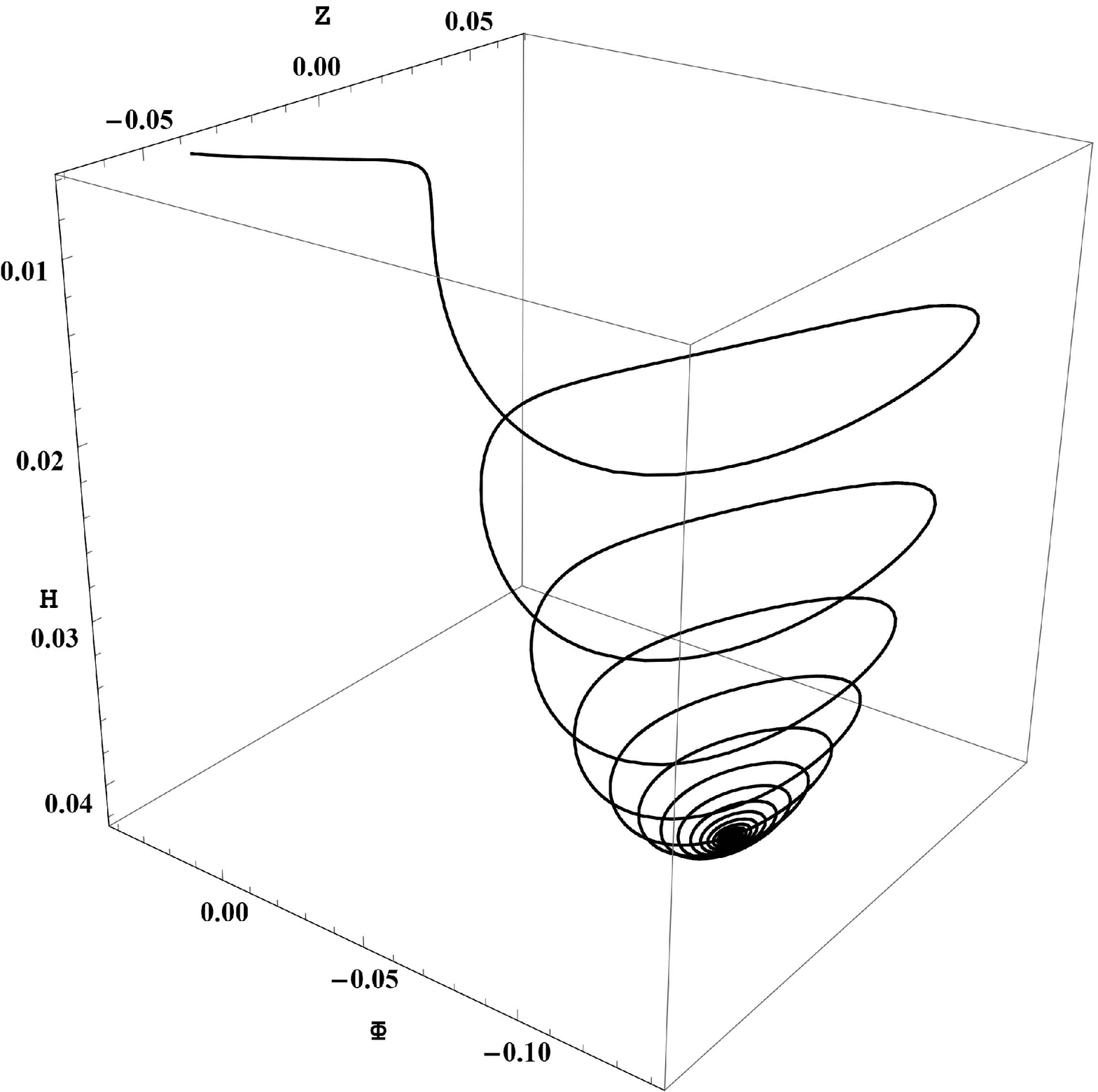}
{A phase portrait of the system for the same case, axes  $\Phi$, $Z$, $h$.\label{fig::3_2_4b}}

Figures \ref{fig::4_2_10} -- \ref{fig::4_2_15} illustrate the results of numerical simulation of the evolution of the system's solution depending on the cosmological constant: $\epsilon_1 = -1$, $\epsilon_2 = -1$, $\alpha_m=-1$, \linebreak $\Lambda_m=\{0, 0.01, 0.1\}$,  $Z(0)=0$, $\Phi(0)=0.1$, $\tau=0 \div 50$.

\TwoFig{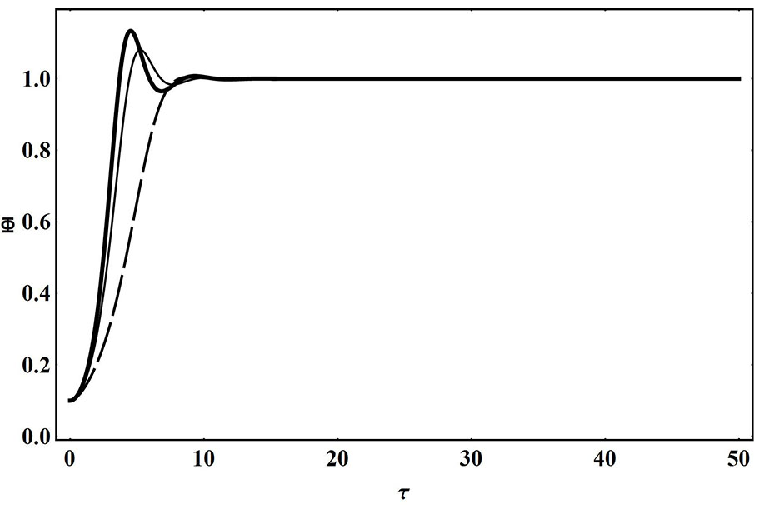}{Evolution of potential of the scalar field $\Phi$ for the case  $\alpha_m=-1$;  $Z(0)=0$; $\tau=0 \div 50$; $\Lambda_m=0$  is a heavy line,  $\Lambda_m=0.01$ is a thin line,$\Lambda_m=0.1$  is a dotted line.\label{fig::4_2_10}}%
{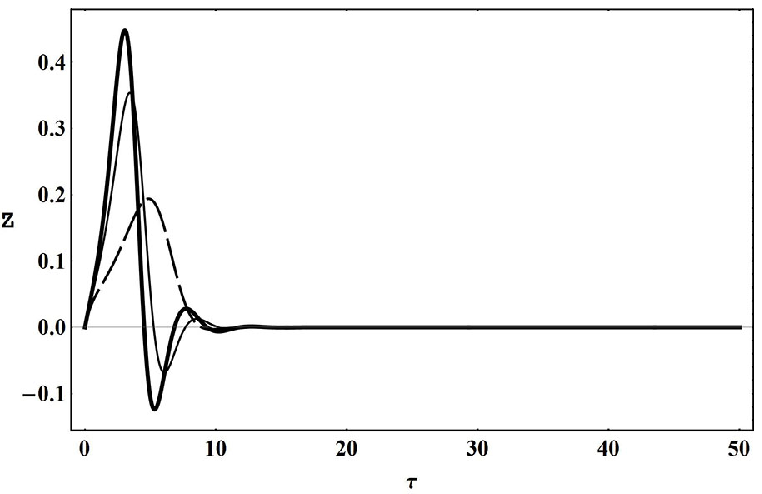}{Evolution of derivative of the scalar field's potential $Z = \dot{\Phi}$.\label{fig::4_2_11}}

\TwoFig{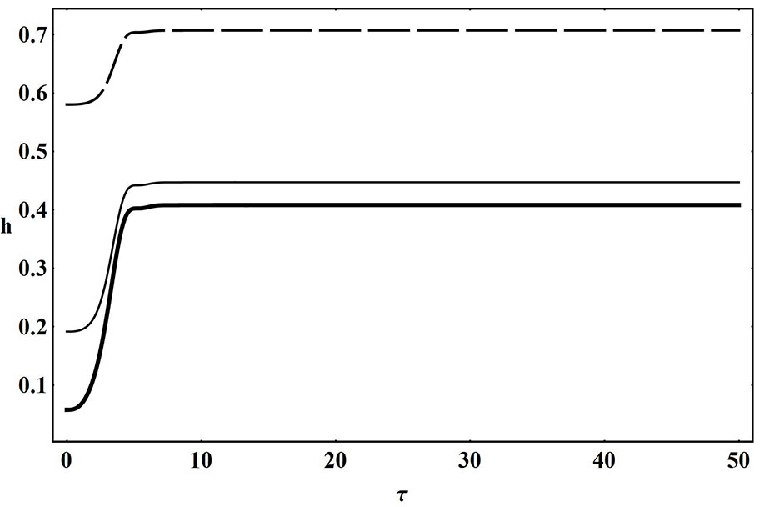}{Evolution of the Hubble constant.  \label{fig::4_2_13}}{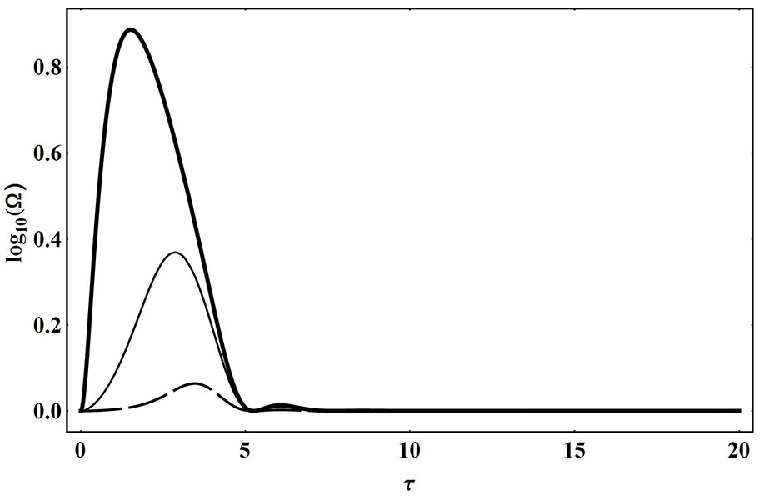}{Evolution of cosmological acceleration's logarithm.  \label{fig::4_2_14}}
\TwoFig{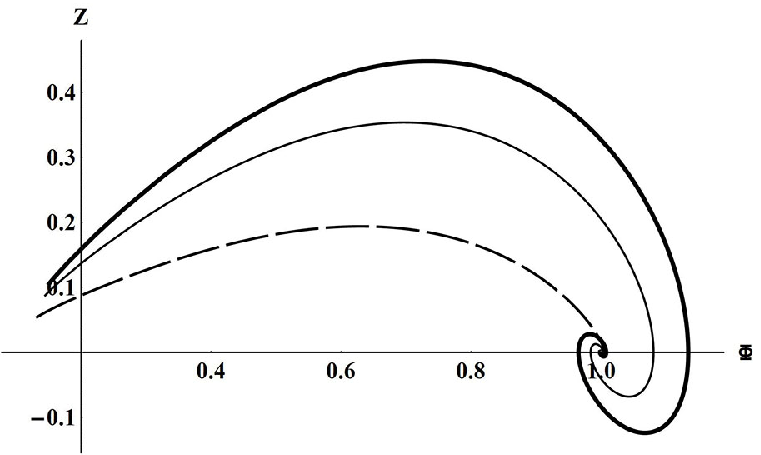}{Phase portrait of the system.  \label{fig::4_2_12}}{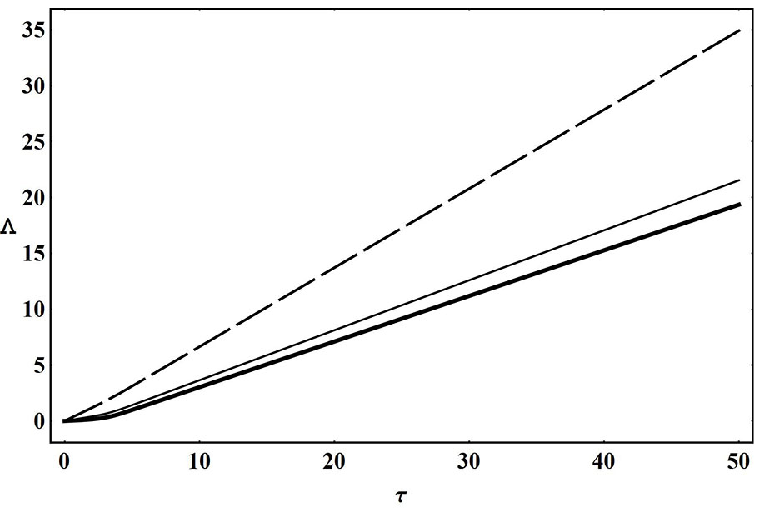}{Evolution of the scale factor's function $\Lambda = \ln a$.  \label{fig::4_2_15}}

\subsection{Numerical Integration of the Model Without a Cosmological Constant ($\Lambda=0$) with a Phantom Scalar Field ($\epsilon_1=-1$, $\epsilon_2=-1$) with a Self-Action ($\alpha_m<0$) }
Figures \ref{fig::4_2_1} -- \ref{fig::4_2_3} show the results of numerical simulation of the system for the following case: $\epsilon_1 = -1$, $\epsilon_2 = -1$, $\alpha_m=-10$, $\Lambda_m=0$,  $Z(0)=-0.4$, $\Phi(0)=\{0.1, 0.05, 0.01\}$, $\tau=0 \div 50$.

\TwoFig{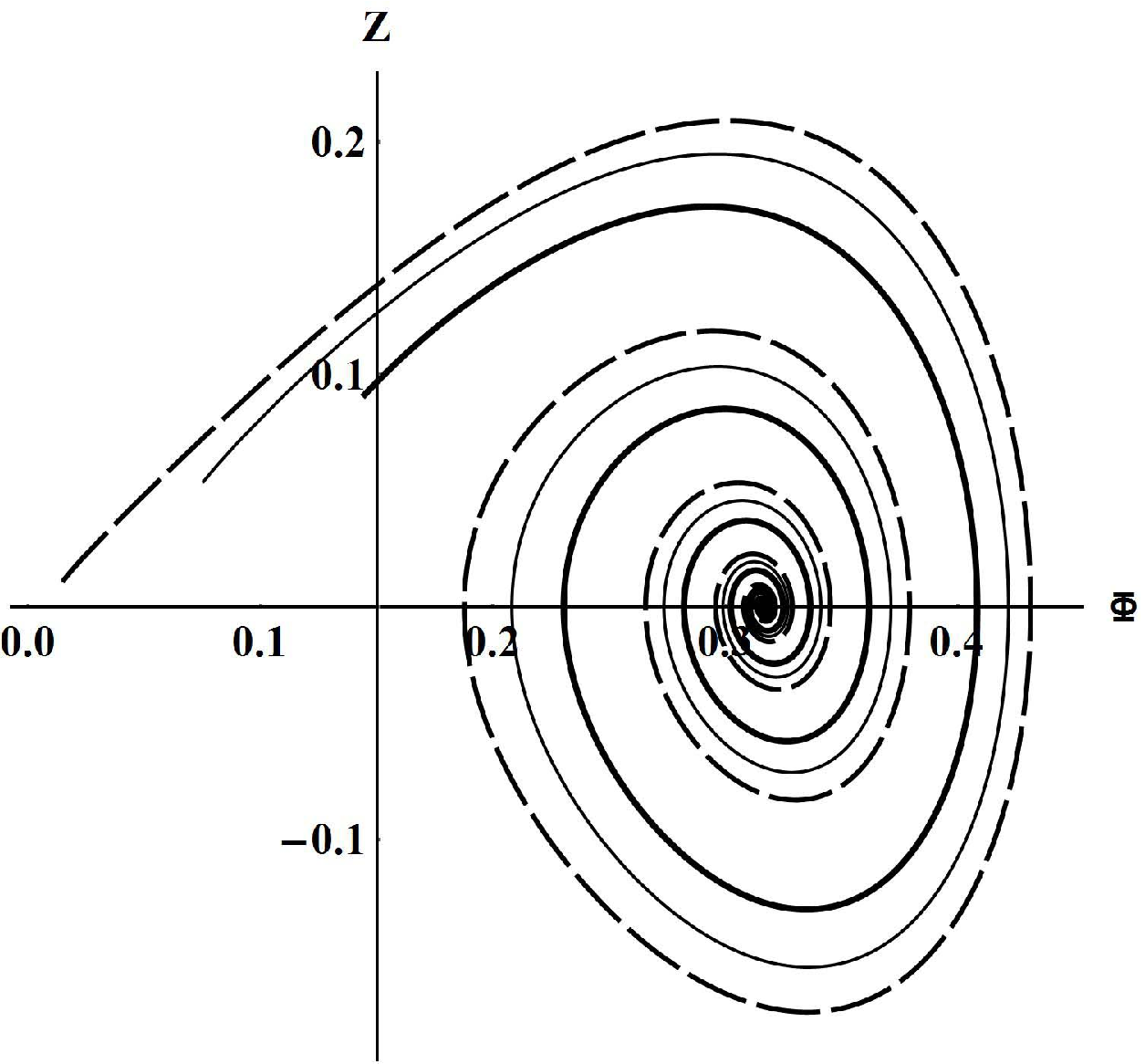}{A phase portrait of the system for the case $\alpha_m=-10$; $\Lambda_m=0$;  $Z(0)=-0.4$; $\tau=0 \div 50$; $\Phi(0)=0.1$ is a heavy line, $\Phi(0)=0.05$ is a thin line, $\Phi(0)=0.01$ is a dotted line.
  \label{fig::4_2_1}}{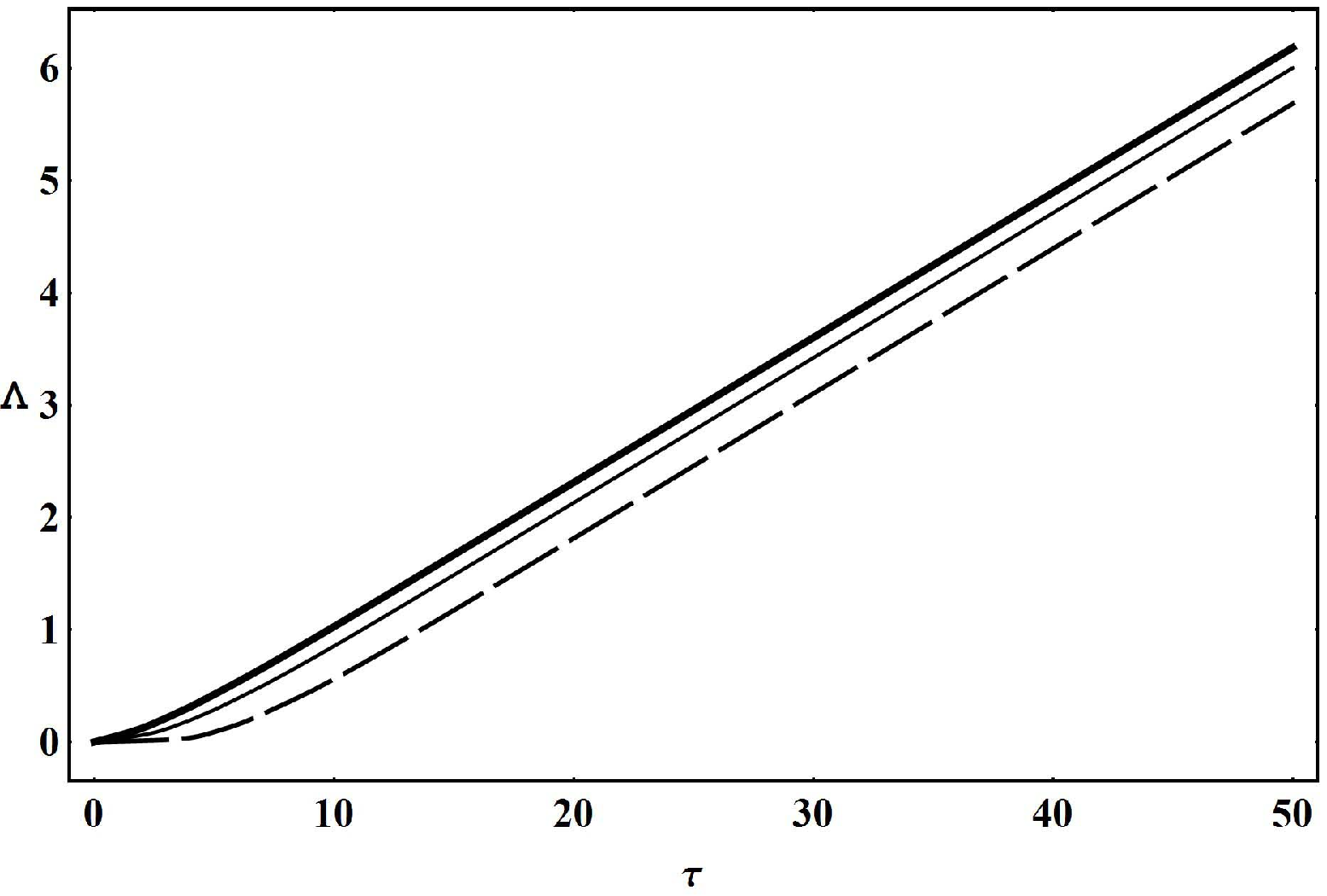}{Evolution of the scale factor's function $\Lambda = \ln a$.  \label{fig::4_2_4}}
\TwoFig{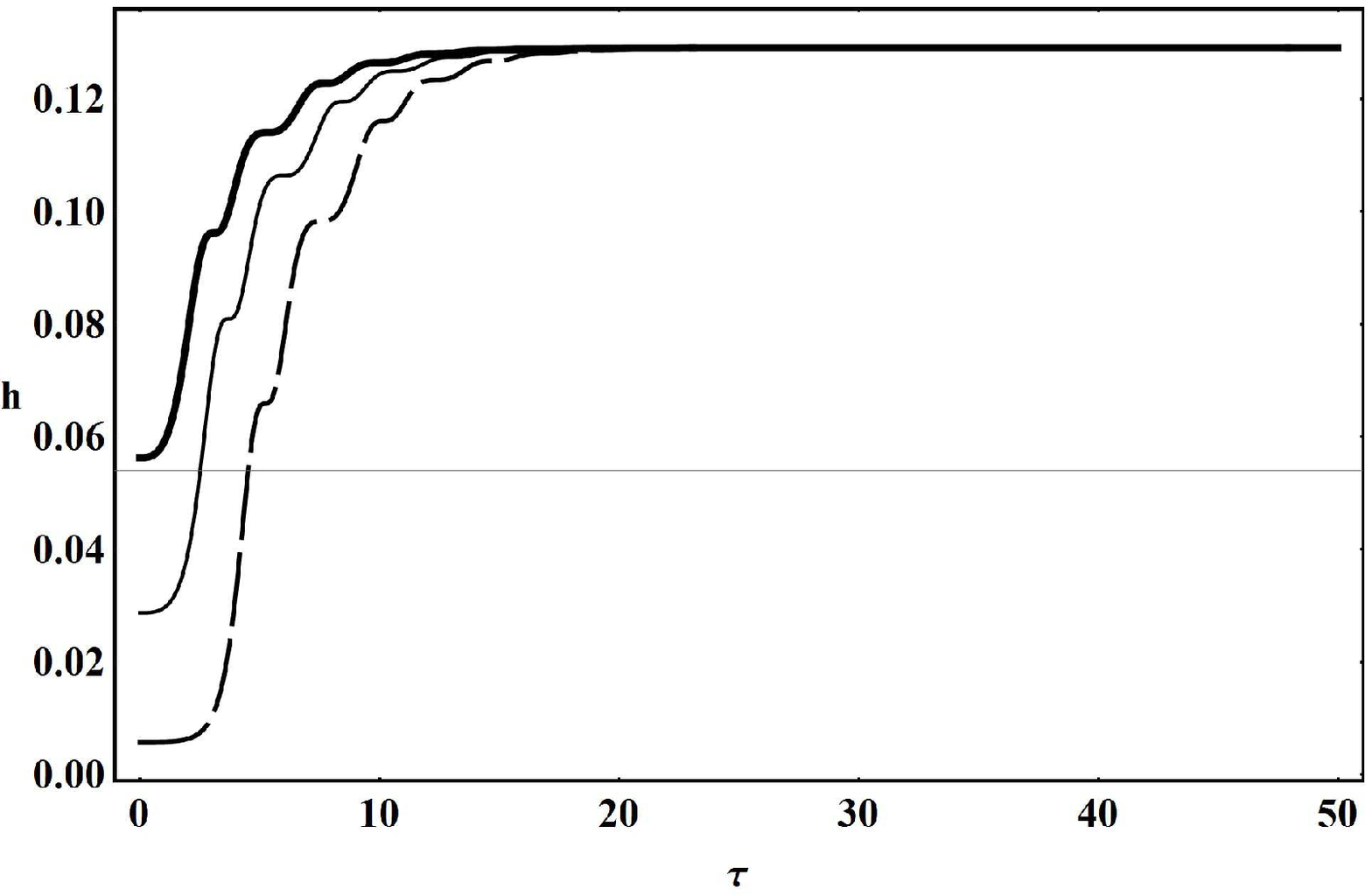}{Evolution of the Hubble constant.  \label{fig::4_2_2}}
{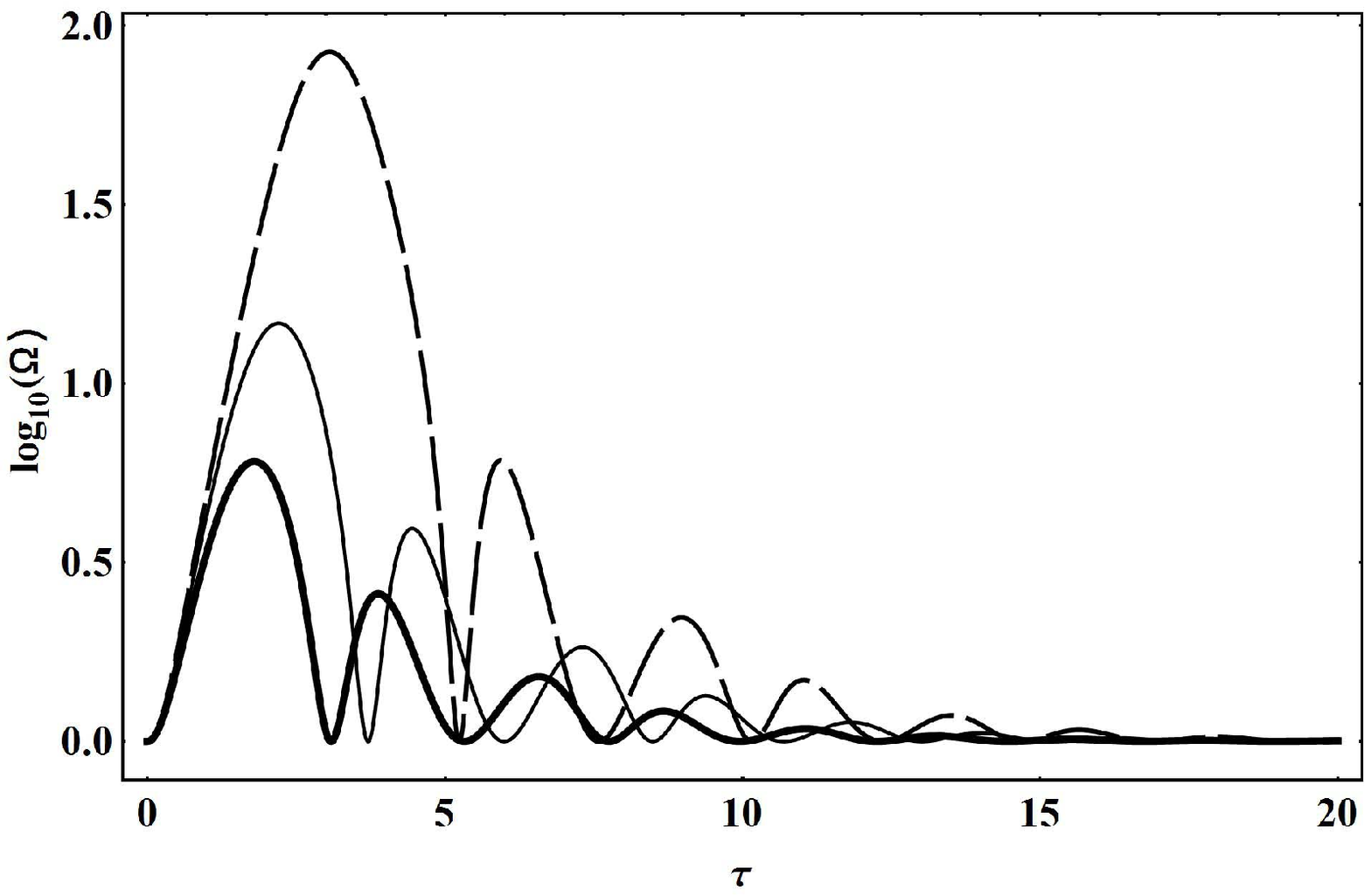}{Evolution of the cosmological acceleration's logarithm.\label{fig::4_2_3}}

Figures \ref{fig::4_2_5} -- \ref{fig::4_2_9} show the results of numerical simulation of the system's evolution depending on the constant of self-action: $\epsilon_1 = -1$, $\epsilon_2 = -1$, $\alpha_m=\{-1, -10, -50\}$, $\Lambda_m=0$,  $Z(0)=0$, $\Phi(0)=0.1$, $\tau=0 \div 50$.

\TwoFig{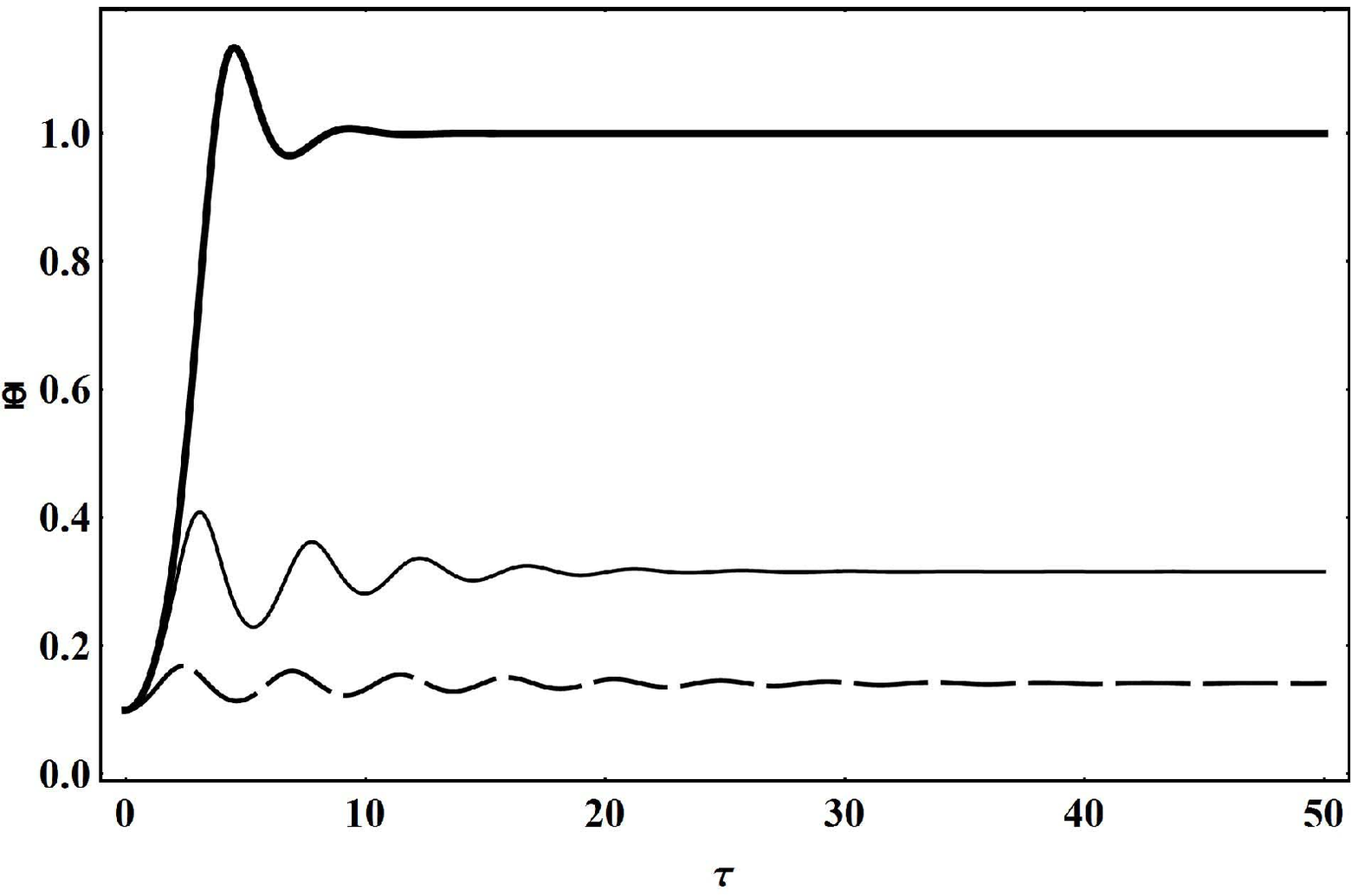}{Evolution of the scalar field's potential $\Phi$ for the case  $\Lambda_m=0$;  $Z(0)=0$; $\tau=0 \div 50$; $\Phi(0)=0.1$; $\alpha_m=-1$ is a heavy line, $\alpha_m=-10$ is a thin line, $\alpha_m=-50$ is a dotted line.
  \label{fig::4_2_5}}{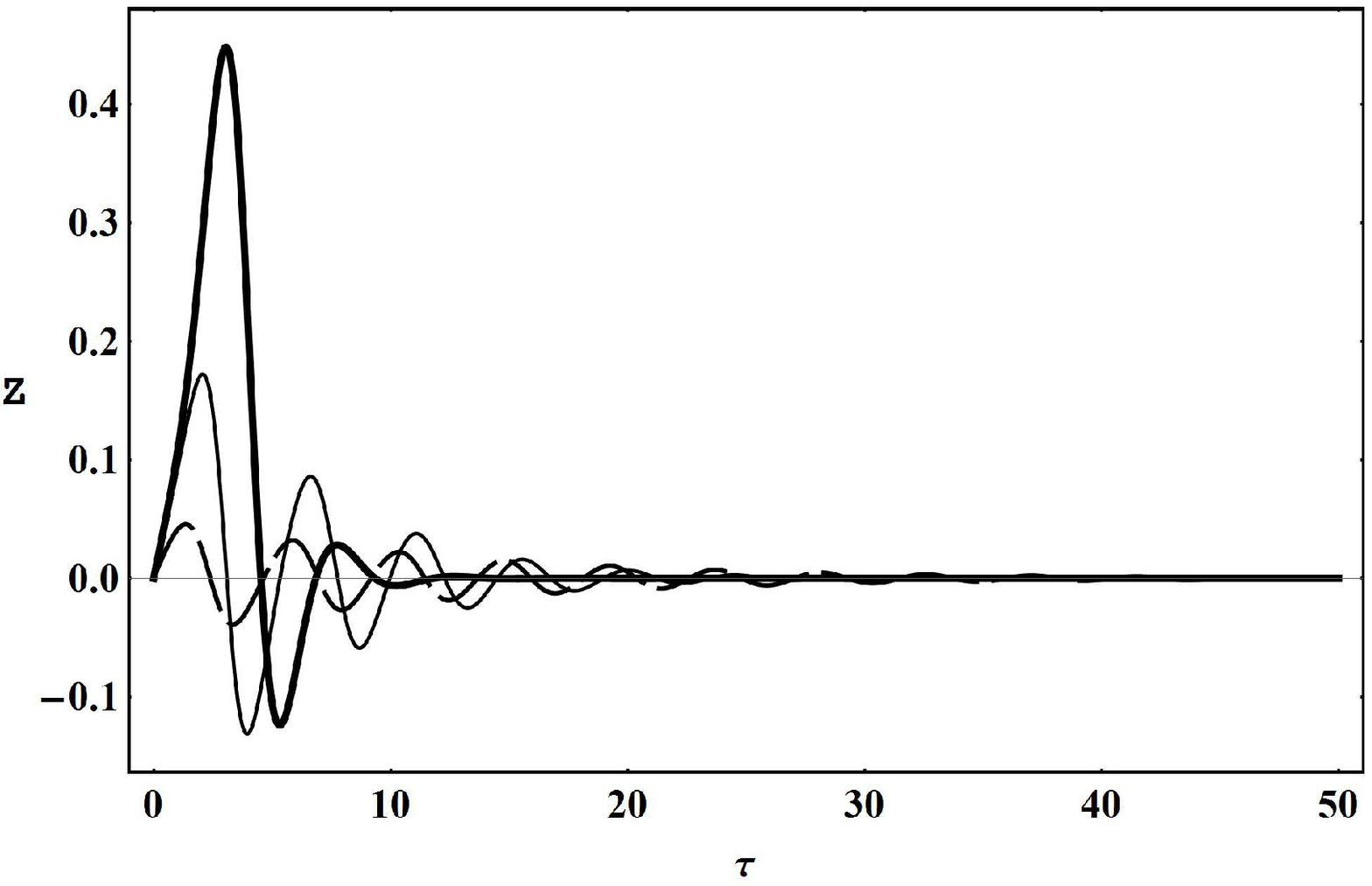}{Evolution of derivative of the scalar field's potential $Z = \dot{\Phi}$.  \label{fig::4_2_6}}
\TwoFig{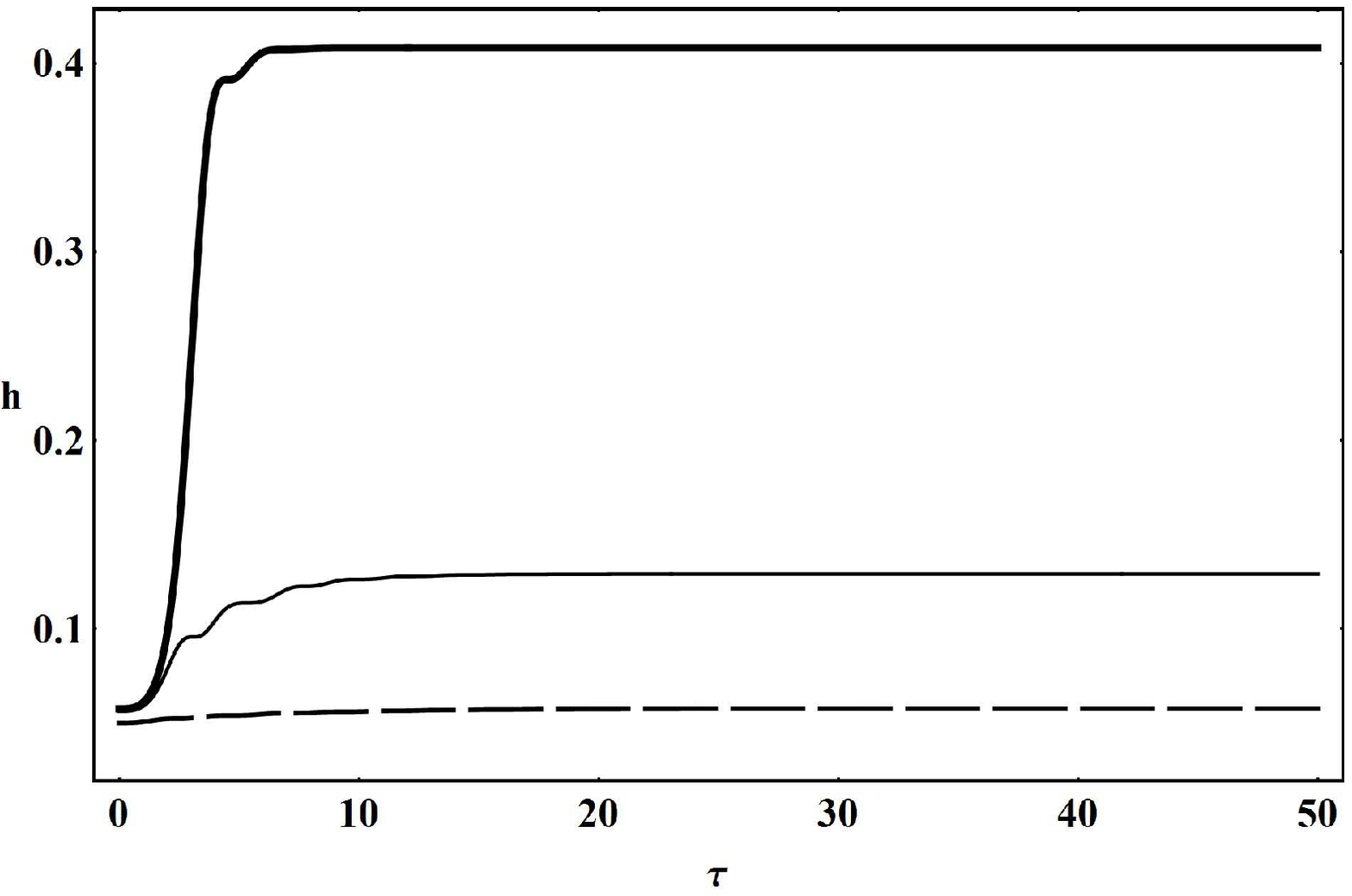}{Evolution of the Hubble constant.  \label{fig::4_2_7}}{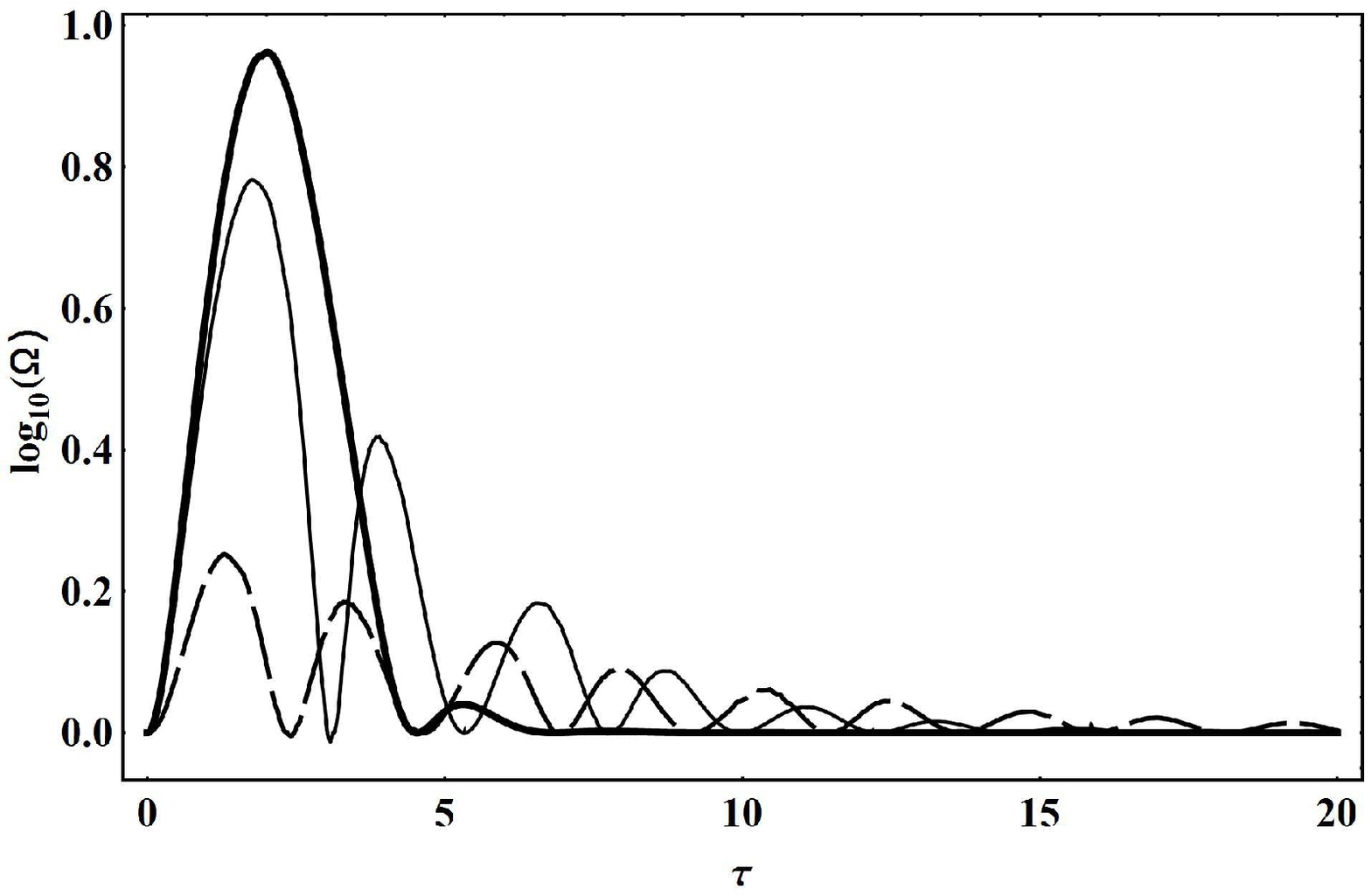}{Evolutino of the cosmological acceleration's logarithm.  \label{fig::4_2_8}}
\Fig{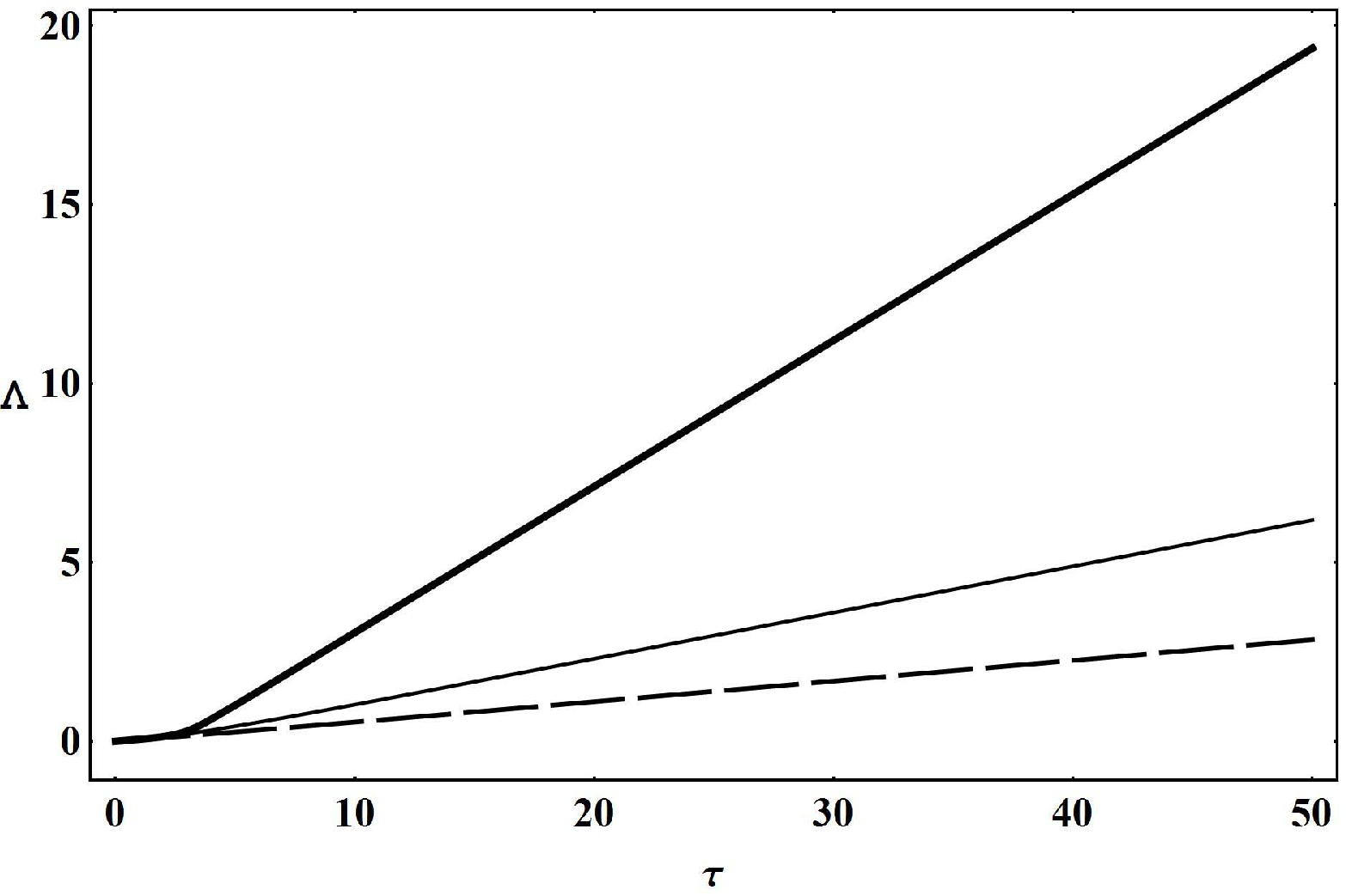}{10}{Evolution of the scale factor's logaritm\linebreak $\Lambda = \ln a$.  \label{fig::4_2_9}}
\subsection{Numerical integration of a Model with a Classical Scalar Field with Higgs Potential of Self-Action.}

Let us consider a classical scalar field with Higgs potential of self-action:

\begin{equation}
V(\Phi) = \frac{\alpha}{2}\left(\Phi^2 - \frac{\beta}{\alpha}\right)^2.
\end{equation}
Thus, the system (\ref{GrindEQ__11_}) can be re-written in the next form:

\begin{eqnarray} \label{higgs}
\Phi ' &=& Z;\nonumber\\
Z' &=& \displaystyle-\sqrt{3} Z\sqrt{ \left(Z^{2} + \frac{\alpha _{m}}{2} \Phi ^{4} - \beta_m \Phi ^{2} + \frac{\beta_m^2}{2\alpha_m} \right) + \Lambda _{m}} \; \displaystyle - \alpha_m \Phi^3 + \beta_{m} \Phi ,
\end{eqnarray}
where it is:
\[\Lambda _{m} \equiv \frac{\Lambda }{m^{2} } ;{\rm \; \; }\alpha _{m} \equiv \frac{\alpha }{m^{2} } ;{\rm \; \; }\beta _{m} \equiv \frac{\beta }{m^{2} } . \]
The system's singular points:
\begin{equation*}
M_0(0,0),\quad M_{\pm}\left(\pm\sqrt{\frac{\beta_m}{\alpha_m}},0\right).
\end{equation*}

Figures \ref{fig::5_2_1} -- \ref{fig::5_2_3} show the results of numerical simulation of the system for the case: $\epsilon_1 = 1$, $\alpha_m=1$, $\beta_m=0.5$, $\Lambda_m=0.000001$,  $Z(0)=-0.1$, $\Phi(0)=1$, $\tau=0 \div 10000$.

\TwoFig{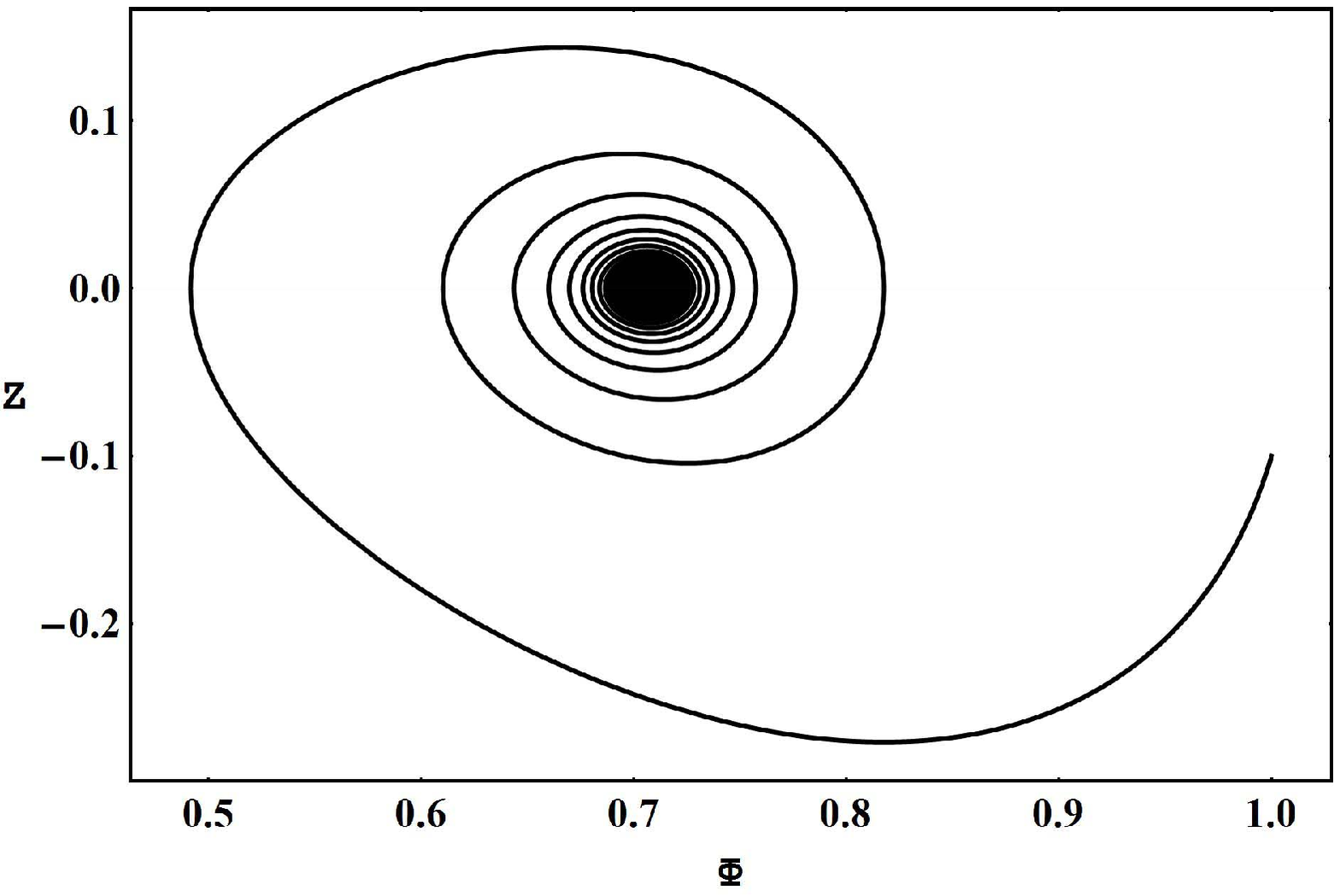}{A phase portrait of the system with Higgs potential of self-action $\tau=0 \div 1000$.  \label{fig::5_2_1}}{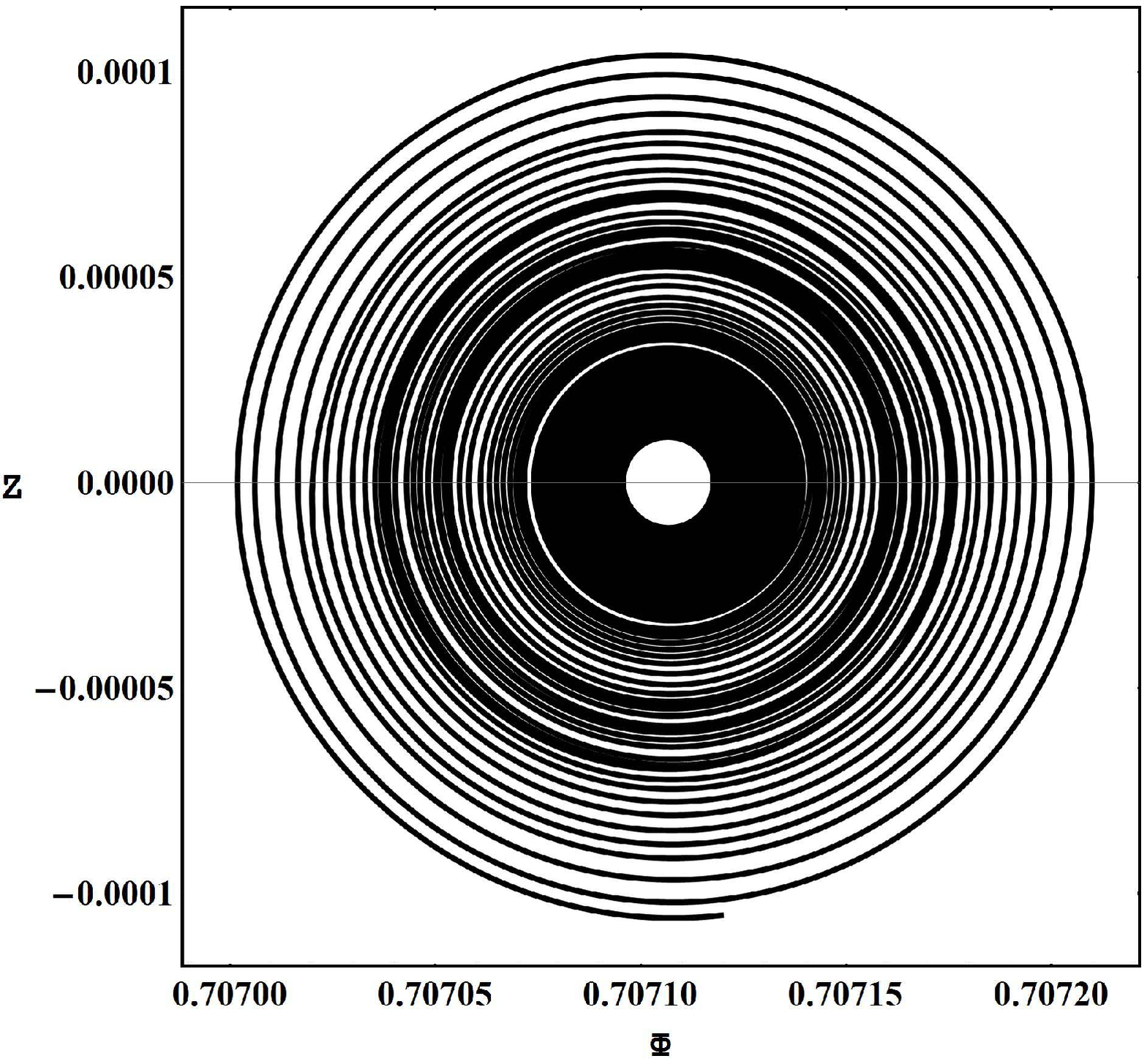}%
{A phase portrait of the system with Higgs potentia of self-action in a large scale:  $\epsilon_1 = 1$, $\alpha_m=1$, $\beta_m=0.5$, $\Lambda_m=0.000001$;  $Z(0)=-0.1$, $\Phi(0)=1$; $\tau=1000 \div 10000$.
  \label{fig::5_2_2}}

\section{Discussion of the Resuts}

In \cite{Zhur_16} it was suggested that in case of classical scalar field at $\Lambda\not=0$ a null singular point should not be a focus and likely it should be corresponded by a limiting cycle corresponding to oscillations of a dynamic system. This consequence was done on the basis of numerical analysis. Let us notice that in this paper we do not consider a second component of the cosmological model which in  \cite{Zhur_16} described a scalarwise neutral ideal flux. Nevertheless, the cited suggestion seems to be incorrect to us, since, from the one hand, numerical analysis of a dynamic system's evolution at increase of the interval of dimensionless time  $\tau$ (up to values of order of $10^5$) shows monotonous decrease of amplitude of the dynamic system's oscillations (Fig. \ref{fig::5_2_5} -- \ref{fig::1_1_3}).

\Fig{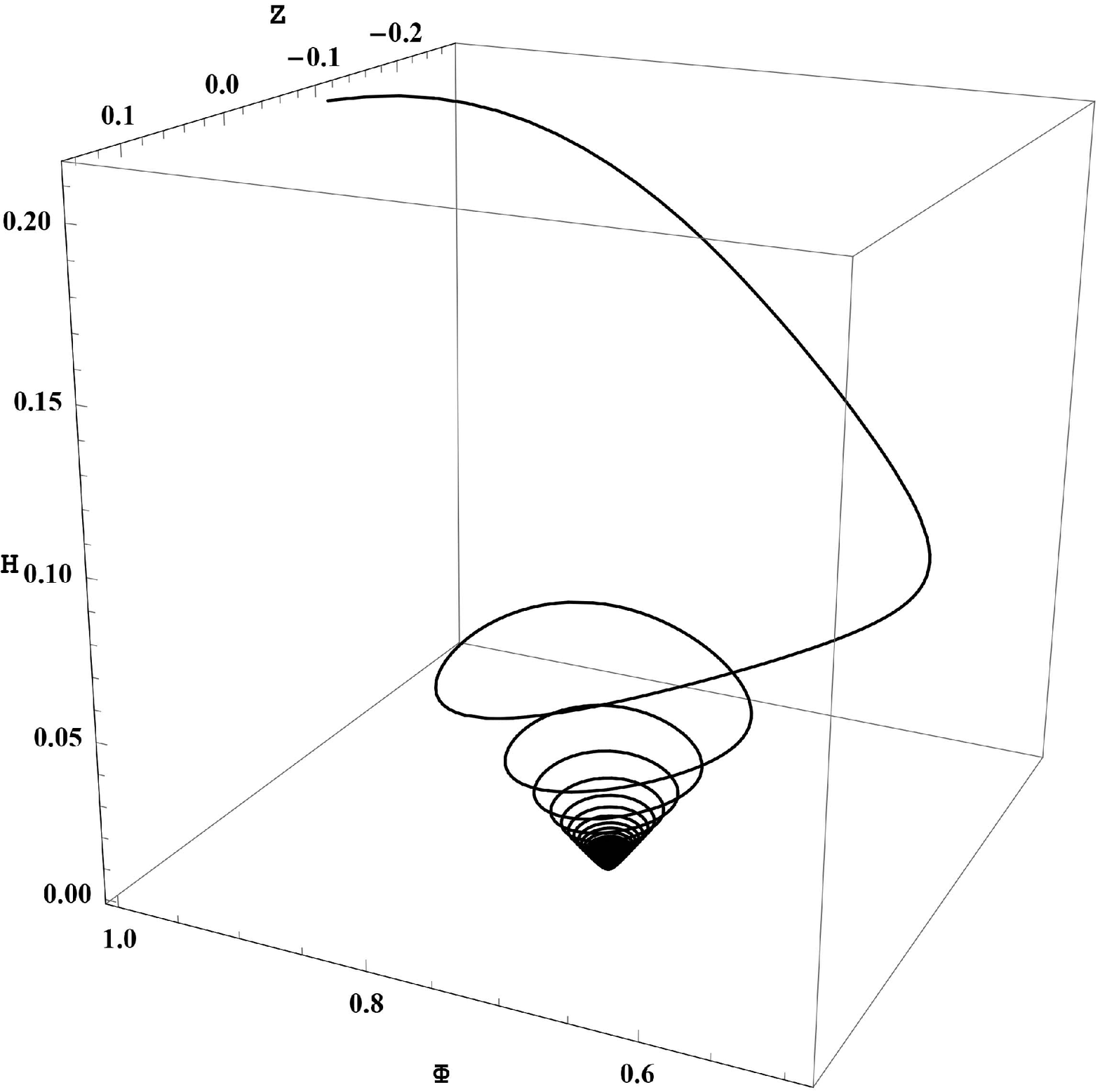}{8}{A 3-dimensional portrait of the system for the same case; axes $\Phi$, $Z$, $h$.  \label{fig::5_2_3}}

From the other hand, supposing at $\tau\to\infty$ $\Phi(\tau)\to0$, $Z(\tau)\to0$ and neglecting the squares of these terms in equations (\ref{GrindEQ__11_}) as compared to $\Lambda_m$, let us reduce the last ones to the following form in case of classical scalar field:
\begin{equation}\label{lin}
\Phi'' +\sqrt{3\Lambda_m}\Phi'+\Phi=0.
\end{equation}
The solution of this equation
\begin{equation}\label{lin_solve}
\Phi=\mathrm{e}^{-\sqrt{\frac{3}{4}\Lambda_m}\tau}\left(C_+\mathrm{e}^{+i\sqrt{1-\frac{3}{4}\Lambda_m}\tau}+C_-\mathrm{e}^{-i\sqrt{1-\frac{3}{4}\Lambda_m}\tau}\ \right);\quad (\Lambda_m<\frac{4}{3})
\end{equation}
describes damped oscillations with a characteristic damping time $\displaystyle\tau_{eff}=\sqrt{\frac{4}{3\Lambda_m}}$. In models without a cosmological term ($\Lambda = 0$) equation (\ref{lin}) describes non-damping oscillations and its solution contains the center \cite{Ignat_16_1_stfi}.

Secondly, the system oscillations near the singular point happen with a period of order:
\begin{equation}\label{T}
T=\frac{2\pi}{\sqrt{1-\frac{3}{4}\Lambda_m}}\sim 2\pi,
\end{equation}
i.e. with a Compton period relative to mass of scalar bosons. These, as was noticed in \cite{Ignat_16_2_stfi} are microscopic times, not available for classical measurements. Therefore it is unjustified to talk about observed oscillations of scale functions ($a(t), H(t), \Omega(t)$) at modern stage of the cosmological evolution. We are most likely talking about continuously damping quantum oscillations of a scalar field which should be interpreted as generation of Higgs bosons which probability decreases exponentially fast with time.


Let us notice, that dynamic equations (\ref{GrindEQ__11_}) are reduced to ordinary differential equation of the following form:
\begin{equation}\label{eq_osc}
\ddot{x}+\beta(x,\dot{x})\dot{x}-V'_x(x)=0,
\end{equation}
where $x(t)\equiv \phi(\tau)$,
\begin{equation}\label{coefs}
\beta(x,\dot{x})= \sqrt{ 2\epsilon_1 \left(\frac{1}{2}\dot{x}^2 -V(x)\right) + \Lambda _{m}}\geq 0.
\end{equation}
Thus, equation (\ref{eq_osc}) physically represents an equation of one-dimensional oscillations in field of Higgs potential $V(x)$ with non-negative coefficient of nonlinear friction $\beta(x,\dot{x})$. Irrespective of explicit form of friction coefficient, the essence of the process, described by the equation (\ref{eq_osc}), is physically transparent --- this is damped oscillations in a potential well $V(x)$. In case of Higgs potential the system descends to one of the stable minimums $V(x)$. Thus, the friction coefficient is a constant at $\Phi\to0, Z\to 0$ and $\Lambda>0$. In case $\alpha=0$ potential $V(x)$ has a form of parabola, therefore the system must descend to its vertex. In case $\Lambda\equiv0$ at $\Phi\to0, Z\to 0$ the friction coefficient tends to null, therefore {\it  non-damped coscillations at $\tau\to\infty$ are only possible exactly in this case}.
Concerning the numerical results and following outcomes of the paper related to existence of limiting cycle of the cosmological dynamic system in case $\Lambda\not\equiv 0$, as can be seen from Fig. \ref{fig::5_2_6} и \ref{fig::5_2_5}, to obtain more precise results on large times of evolution it is required to apply more precise methods of integration of nonlinear equations.

\TwoFig{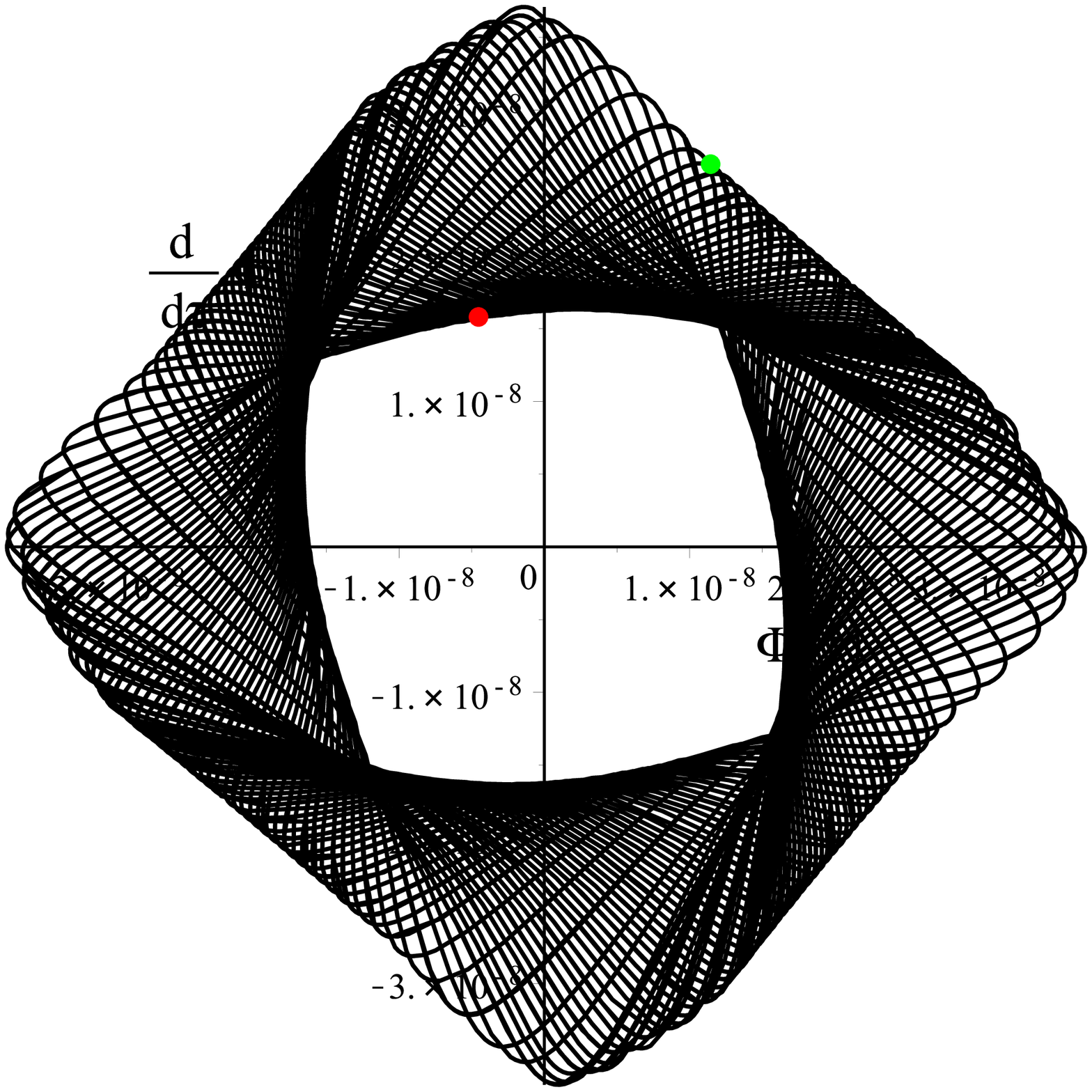}{A phase portrait of the system \eqref{GrindEQ__11_} in a large scale for the case $\alpha_m=0$; $\Lambda_m=0.001$; $\tau=9000 \div 10000$; $\Phi(0)=1;Z(0)=0$. The plot was obtained using the standard Runge - Kutta method of 4-5 orders in Maple 18.\label{fig::5_2_6}}{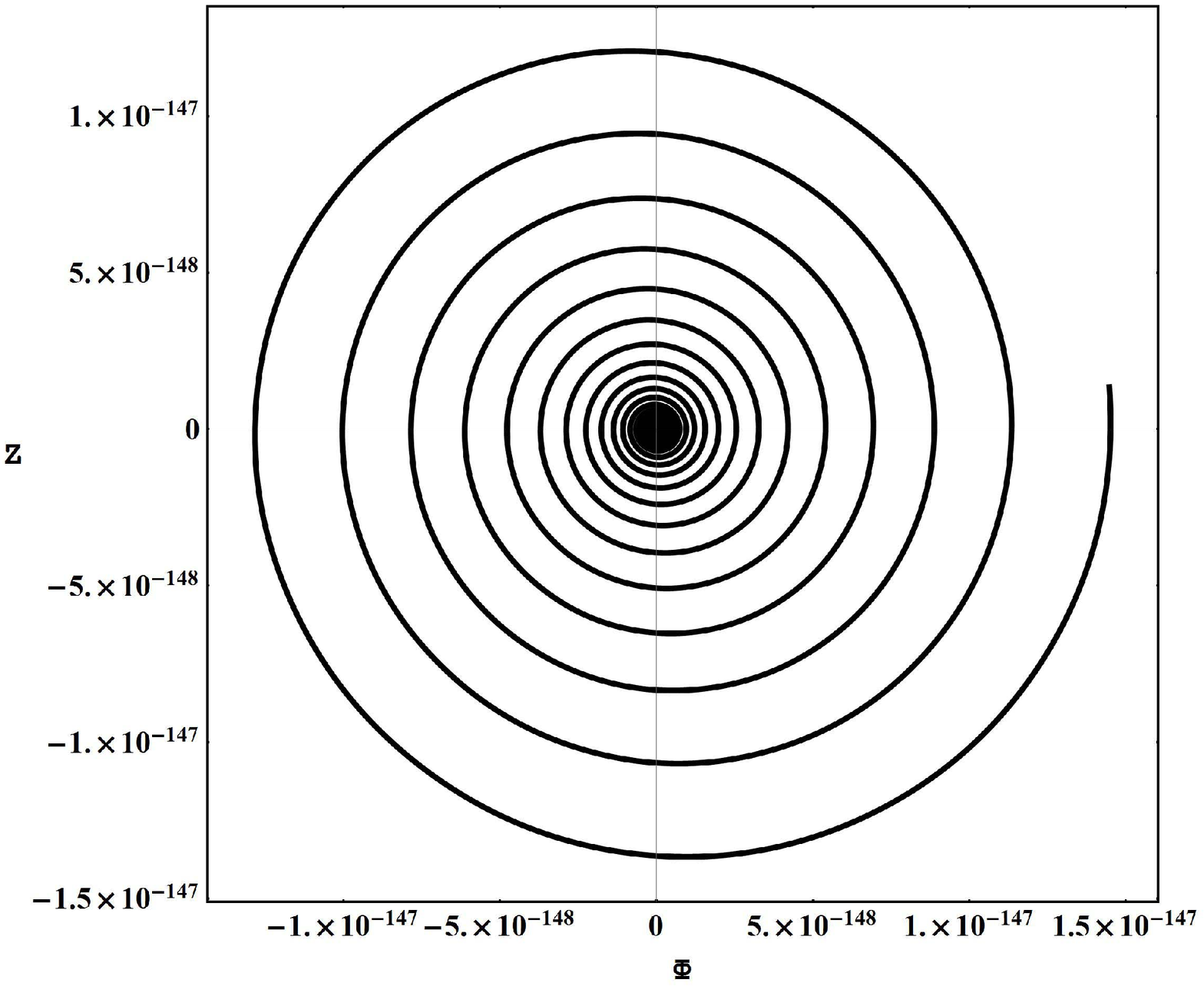}{A phase portrait of the system \eqref{GrindEQ__11_} in a large scale for the case $\alpha_m=0$; $\Lambda_m=0.001$; $\tau=9000 \div 10000$; $\Phi(0)=1;Z(0)=0$. The plot was obtained using the Runge - Kutta integration method of improved accuracy of 7-8 orders in Maple 18.\label{fig::5_2_5}}
From the comparison of graphs on Fig. \ref{fig::5_2_6} and \ref{fig::5_2_5} it is seen that the Runge - Kutta method of 4-5 orders at great values of the dynamic system's evolution and small values of scalar field's potential and its derivative leads to essential errors. The phase diagram shown on Fig. \ref{fig::5_2_6} is also very interesting however it does not represent the facts. We can see a correct phase diagram on Fig. \ref{fig::5_2_5} which was obtained for the same case by Runge - Kutta integration method of improved accuracy of 7-8 orders. Exactly this diagram describes damed oscillations  (\ref{lin_solve}). Let us pay attention to remarkable accuracy of calculations which used Runge - Kutta method of 7-8 orders: the spiral's radius on the considered stage is  $10^{-192}$! Let us also pay attention to the next circumstance: phase diagram on Fig. \ref{fig::5_2_6}, obtained using Runge - Kutta method of 4-5 orders in this case gives us the values of the potential and its derivative, overevaluated by 184 orders!!

Fig.  \ref{fig::5_2_7} --- \ref{fig::5_2_10} show phase diagrams of the dynamic system at large times on case of null value of the cosmological constant. As is seen from the represented phase diagrams, phase diagrams in case $\Lambda=0$ {\it are very similar to} limiting cycles with almost strictly circular orbit in our variables. However, one can also notice a firm tendency for decrease of these cycles' radius with growth of evolution time of a dynamic system.

Tendency of phase trajectories evolution is so that
\begin{equation}\label{Phi0(tau)}
\Phi^2(\tau)+\Phi'\ \!\! ^2(\tau)=\Phi^2_0(\tau)\approx \mathrm{const};\quad (\tau\gg 1),
\end{equation}
where  the approximate relationship $\Phi_0(\tau)\varpropto \tau^{-5/2}$  fulfills to a certain instant of time and represents a slowly changing function as compared with a phase of oscillations; after this $\Phi_0(\tau)$ starts to fall quite much faster.

The general tendency for decrease of the oscillations amplitude in case $\Lambda=0, \alpha=0$ can be understood also on the basis of analysis of the field equation
\begin{equation}\label{Phi''Eq}
\Phi''+\sqrt{3}\Phi'\sqrt{\Phi^2+\Phi'\ \!\!^2}+\Phi=0
\end{equation}
taking into account properties (\ref{Phi0(tau)}. Let us notice a curious fact: functions
\begin{equation}\label{exact_sol}
\Phi_\pm =\mathrm{C}_\pm \mathrm{e}^{\pm i\tau},
\end{equation}
where $\mathrm{C}_\pm$ are arbitrary constants, are the exact solutions of the field equation (\ref{Phi''Eq}). Actually:
\begin{equation}\label{equiv}
\Phi''_\pm+\Phi_\pm\equiv0; \quad \Phi^2_\pm+\Phi'\ \!\!^2_\pm\equiv0.
\end{equation}
However, it is obvious that no real solutions will turn to null the radicand in the field equation (\ref{Phi''Eq}).

Supposing
\begin{equation}\label{Phi->phi}
\Phi(\tau)=\frac{1}{\sqrt{3}}\phi(\tau),
\end{equation}
let us reduce the field equation (\ref{Phi''Eq}) to the next form:
\begin{equation}\label{phi''Eq}
\phi''+\phi'\sqrt{\phi^2+\phi'\ \!\!^2}+\phi=0.
\end{equation}

Actually, taking into account (\ref{Phi0(tau)} and supposing that in (\ref{Phi''Eq}) it is:
$$\Phi=\Phi_0(\tau)\mathrm{e}^{i\tau},$$
we find:
\begin{equation}\label{WKB}
\Phi''_0+2i\Phi'_0+\sqrt{3}\Phi'_0\Phi_0+i\sqrt{3}\Phi^2_0=0.
\end{equation}
Accounting $\Phi'_0\ll \Phi_0$, let us at a first approximation obtain an amplitude evolution law from (\ref{WKB}):
\begin{equation}
\Phi_0\varpropto \frac{1}{\sqrt{3}\tau},
\end{equation}
which confirms the conclusion about decrease of the oscillations' amplitude with time.

\TwoFig{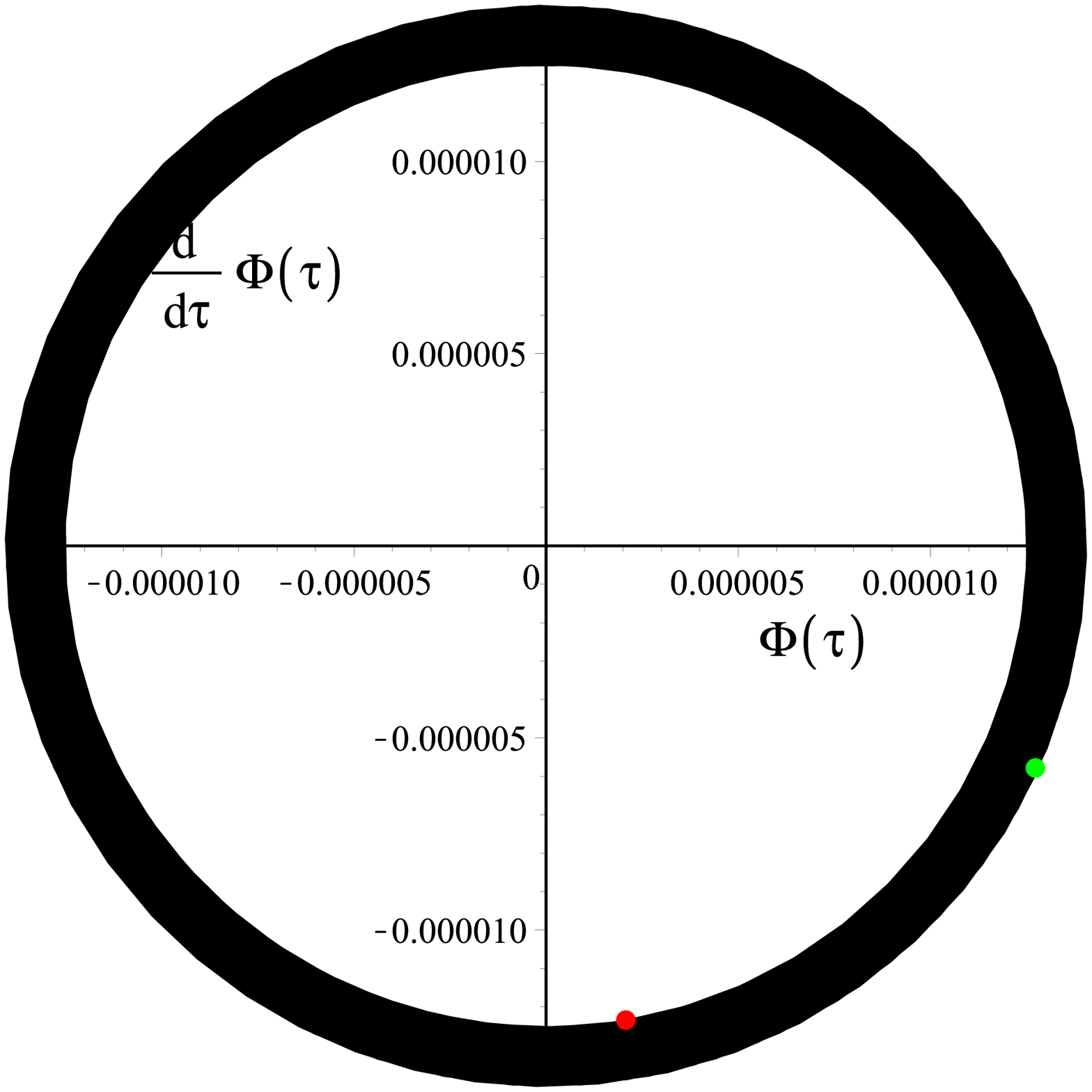}{A phase portrait of the system \eqref{GrindEQ__11_} in a large scale for the case $\alpha_m=0$; $\Lambda_m=0$; $\tau=9000 \div 10000$; $\Phi(0)=1;Z(0)=0$. This plot is obtained using the standard Runge - Kutta method of 4-5 orders in Maple 18.\label{fig::5_2_7}}{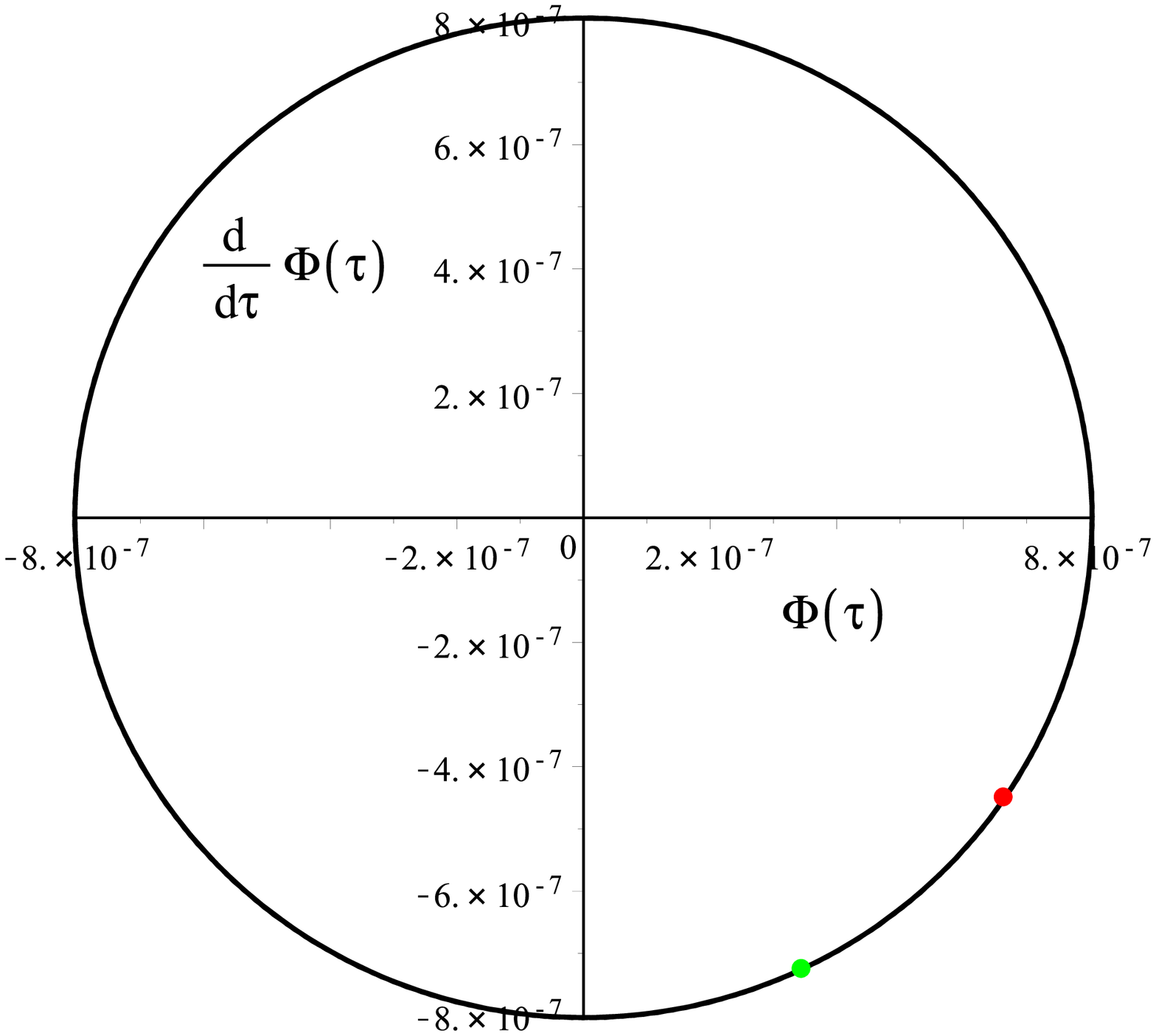}{A phase portrait of the system  \eqref{GrindEQ__11_} in a large scale for the case $\alpha_m=0$; $\Lambda_m=0$; $\tau=100000 \div 100100$; $\Phi(0)=1;Z(0)=0$. This plot is obtained using the high accuracy Runge - Kutta method of 7-8 orders in Maple 18.\label{fig::5_2_8}}
\TwoFig{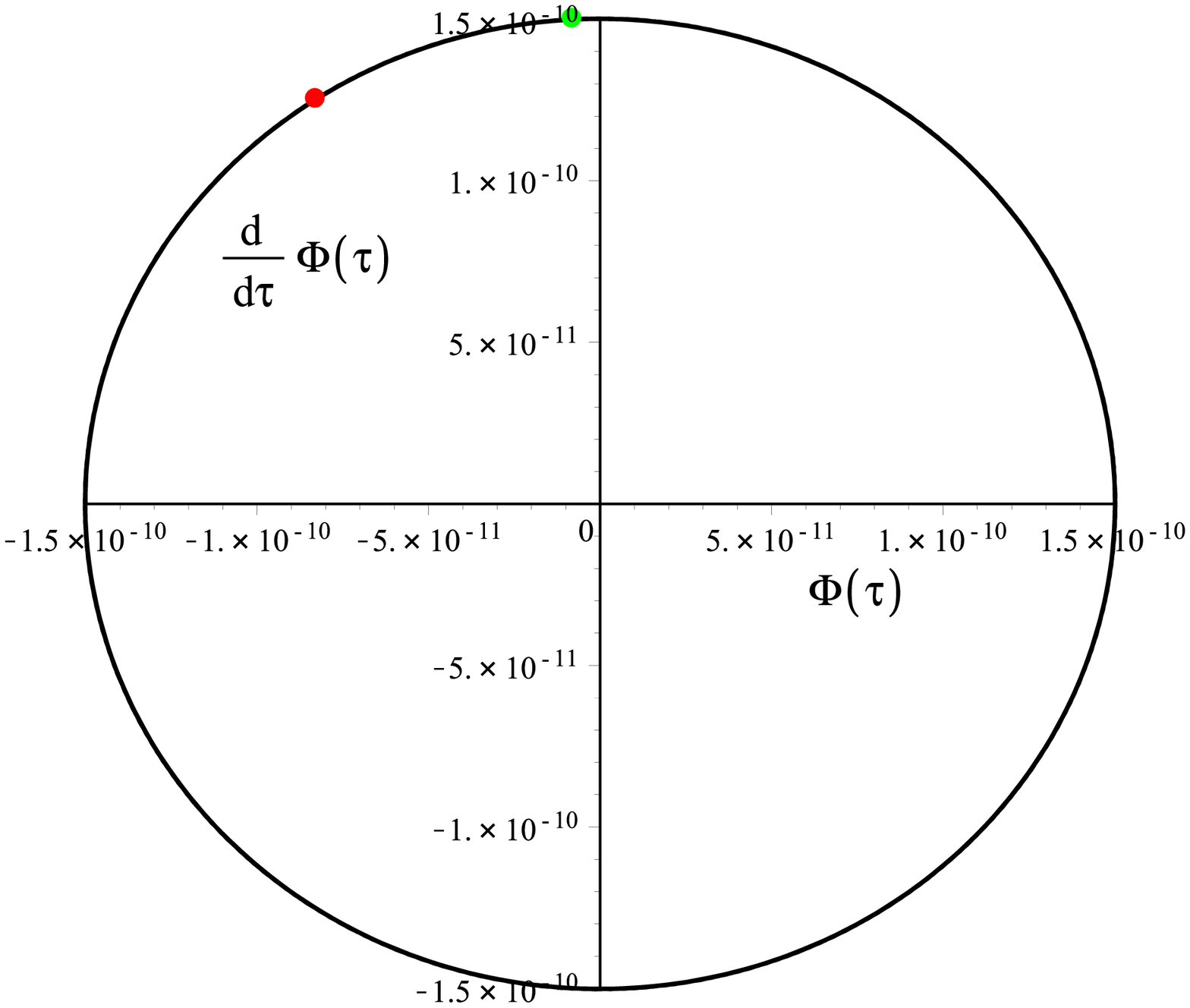}{A phase portrait of the system \eqref{GrindEQ__11_} in a large scale for the case $\alpha_m=0$; $\Lambda_m=0$; $\tau=1000000 \div 1000100$;; $\Phi(0)=1;Z(0)=0$.  This plot is obtained using the standard Runge - Kutta method of 4-5 orders in Maple 18.\label{fig::5_2_9}}{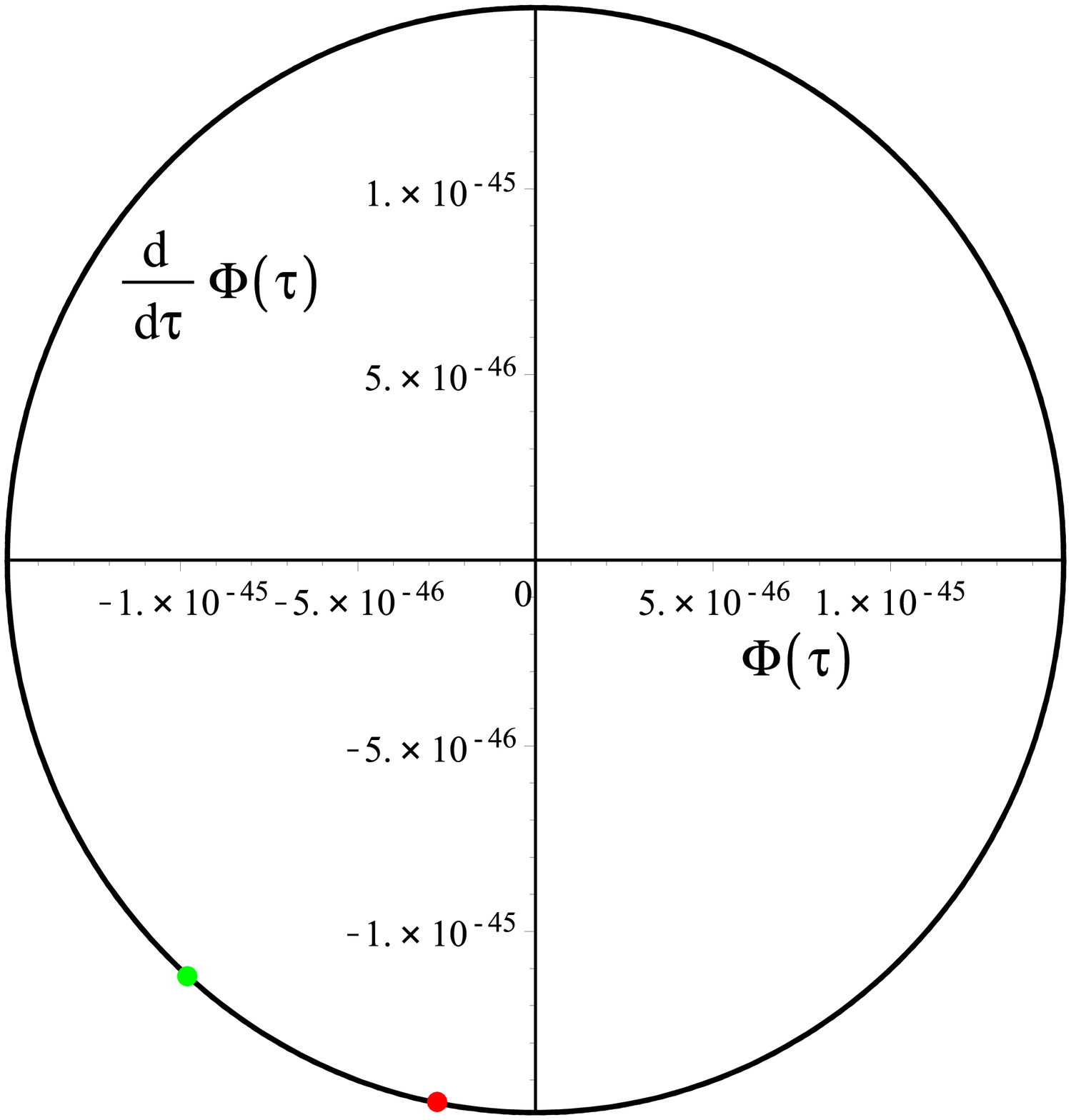}{A phase portrait of the system \eqref{GrindEQ__11_} in a large scale for the case $\alpha_m=0$; $\Lambda_m=0$; $\tau=10000000 \div 10000100$; $\Phi(0)=1;Z(0)=0$. This plot is obtained using the high accuracy Runge - Kutta method of 7-8 orders in Maple 18.\label{fig::5_2_10}}

\begin{flushleft}
{\bf{Acknowledgements}}
\end{flushleft}

In conclusion, the Authors express their gratitude to the members of MW seminar for relativistic kinetics and cosmology of Kazan Federal University for helpful discussion of the work.

\vspace{10pt}
\small

\makeatletter
\@addtoreset{equation}{section}
\@addtoreset{footnote}{section}
\renewcommand{\section}{\@startsection{section}{1}{0pt}{1.3ex
plus 1ex minus 1ex}{1.3ex plus .1ex}{}}

{ 

\renewcommand{\refname}{{\rm\centerline{СПИСОК ЛИТЕРАТУРЫ}}}

\end{document}